%% file: ppan_tifs.tex
\documentclass[journal]{IEEEtran}
\pdfoutput=1
\usepackage{graphicx}
\graphicspath{{Figures/}}
\DeclareGraphicsExtensions{.pdf, .png, .jpg}

\usepackage[dvipsnames]{xcolor}
\usepackage{subcaption}
\usepackage{url}            % simple URL typesetting
\usepackage{booktabs}       % professional-quality tables
\usepackage{nicefrac}       % compact symbols for 1/2, etc.
\usepackage{bm}             % bold math
\usepackage{amssymb, amsmath, amsthm, amsfonts}

\DeclareMathOperator{\diag}{diag}
\DeclareMathOperator{\KL}{KL}
\newtheorem{propn}{Proposition}
\usepackage{mathtools, array}
\newcolumntype{P}[1]{>{\centering\arraybackslash}p{#1}}
\usepackage{hyperref}       % hyperlinks
\newcommand\independent{\protect\mathpalette{\protect\independenT}{\perp}}
\def\independenT#1#2{\mathrel{\rlap{$#1#2$}\mkern2mu{#1#2}}}
\def\edits#1{{\color{black} #1}}

\title{Privacy-Preserving Adversarial Networks}

\author{Ardhendu Tripathy,
Ye Wang, and
Prakash Ishwar
\thanks{A.~Tripathy is with University of Wisconsin-Madison, WI 53703, email:~{\tt astripathy@wisc.edu}, and performed a part of this work during an internship at MERL.}
\thanks{Y.~Wang is with Mitsubishi Electric Research Laboratories (MERL), Cambridge, MA 02139, email:~{\tt yewang@merl.com}.}
\thanks{P.~Ishwar is with Boston University, Boston, MA 02215, email:~{\tt pi@bu.edu}.}
}

\begin{document}

\maketitle

\begin{abstract}
We propose a data-driven framework for optimizing privacy-preserving data release mechanisms to attain the information-theoretically optimal tradeoff between minimizing distortion of useful data and concealing \edits{specific} sensitive information. Our approach employs adversarially-trained neural networks to implement randomized mechanisms and to perform a variational approximation of mutual information privacy. We validate our Privacy-Preserving Adversarial Networks (PPAN) framework via \edits{proof-of-concept} experiments on discrete and continuous synthetic data, as well as the MNIST handwritten digits dataset. For synthetic data, our model-agnostic PPAN approach achieves tradeoff points very close to the optimal tradeoffs that are analytically-derived from model knowledge. In experiments with the MNIST data, we visually demonstrate a learned tradeoff between minimizing the pixel-level distortion versus concealing the written digit.
\end{abstract}

\IEEEpeerreviewmaketitle

\input{intro}
\input{framework}
\input{results}

\input{cont-results}
\input{mnist}

\input{analytical_PU}
\input{conclusion}
%\newpage
\bibliography{refs}
\bibliographystyle{abbrv}

%\newpage
\appendices

\input{algorithm}

\input{appendix}

\input{side_info.tex}
\input{MI_utility}
\input{surrogate.tex}
\input{GMM}
\input{discrete_baseline}
\input{scatter_viz}
\end{document}

%% file: intro.tex
\section{Introduction}

Our work addresses the problem of privacy-preserving data release,
where the goal is to release useful data while also limiting the exposure of associated sensitive information.
Approaches that involve data modification must consider the tradeoff between
concealing sensitive information and minimizing distortion to preserve data utility.
However, practical optimization of this tradeoff can be challenging when we wish to quantify privacy via statistical measures (such as mutual information) and the actual statistical distributions of data are unknown.
In this paper, we propose a data-driven framework involving adversarially trained neural networks to design privacy-preserving data release mechanisms that approach the \edits{information-theoretically optimal} privacy-utility tradeoffs.

Privacy-preserving data release is a broad and widely explored field, where the study of principled methods have been well motivated by highly publicized leaks stemming from the inadequacy of simple anonymization techniques, such as reported in~\cite{Sweeney-CMU2000-SimplyIdentify, NarayananS-SP08-DeAnonNetlix}.
A wide variety of methods to statistically quantify and address privacy have been proposed,
such as $k$-anonymity~\cite{Sweeney02-kAnon},
$L$-diversity~\cite{MachanavajjhalaKGV-TKDD07-Ldiversity},
$t$-closeness~\cite{LiLV-07-tCloseness},
and differential privacy~\cite{DworkMNS-TOC06-DiffPriv}.
In our work, we focus on an information-theoretic approach where privacy is quantified by the mutual information between the data release and the sensitive information~\cite{Yamamoto-IT83-RateDistortionSecrecy, RebolloFD-TKDE10-tCloseToPRAMviaIT, CalmonF-Allerton12, SankarRP-TIFS13-UtilityPrivacy, ITA2016}.

Unlike the \edits{other privacy measures} mentioned earlier, mutual information \edits{depends specifically on the} statistical distribution of the data.
\edits{Requiring consideration of the data distribution is a practical hindrance, 
however measuring privacy while} ignoring the data distribution altogether can weaken the scope of privacy guarantees.
For example, an adversary armed with only mild knowledge about the correlation of the data\footnote{Note that even when data samples are inherently independent, the prior knowledge of an adversary could become correlated when conditioned on particular side information.}
can undermine the practical privacy protection of differential privacy, as noted in examples given by~\cite{KiferM-2011-NoFreeLunch, CalmonF-Allerton12, LiuCM-2016-DDP, wang2017privacy}.
While model assumptions are avoided in the definition of differential privacy, independence across individuals in the dataset is implicitly required to avoid undermining privacy guarantees~\cite{KiferM-2011-NoFreeLunch}.
The example in~\cite[Sec.~V]{CalmonF-Allerton12} demonstrates that an $\epsilon$-differentially private mechanism can leak 
%an unbounded amount of 
sensitive information on the order of $O(\epsilon^2 \log n)$, \edits{in terms of mutual information}, where $n$ is size of the dataset.
Moreover, differential privacy does not satisfy the so-called \textit{linkage inequality}~\cite[Def.~2]{wang2017privacy}, which \edits{captures the notion that privacy guarantees should also limit the disclosure of other sensitive information linked to the primary data considered, as explained further in~\cite{wang2017privacy}.}
%is a basic property expected of privacy measures as argued in~\cite{wang2017privacy}.
\edits{While settling the debate over which privacy measure is most appropriate is beyond the scope of this paper, we nonetheless focus on mutual information privacy, 
%while providing 
and develop a data-driven approach that addresses the practical drawback of requiring distributional knowledge 
for mutual information privacy.}
%We therefore focus on mutual information as the privacy measure in this work.

We build upon the non-asymptotic, information-theoretic framework introduced by~\cite{RebolloFD-TKDE10-tCloseToPRAMviaIT, CalmonF-Allerton12},
where the sensitive and useful data are respectively modeled as random variables $X$ and $Y$.
We also adopt the extension considered in~\cite{ITA2016}, where only a (potentially partial and/or noisy) observation $W$ of the data is available.
In this framework, the design of the privacy-preserving mechanism to release $Z$ is formulated as the optimization of the tradeoff between minimizing privacy-leakage quantified by the mutual information $I(X; Z)$ and minimizing an expected distortion $\mathbb{E}[d(Y, Z)]$.
This non-asymptotic framework has strong connections to generalized rate-distortion problems (see discussion in~\cite{RebolloFD-TKDE10-tCloseToPRAMviaIT, CalmonF-Allerton12, wang2017privacy}),
as well as related asymptotic privacy frameworks where communication efficiency is also considered in a rate-distortion-privacy tradeoff~\cite{Yamamoto-IT83-RateDistortionSecrecy, SankarRP-TIFS13-UtilityPrivacy}.

In principle, when the data distribution is known, the optimal design of the privacy-preserving mechanism can be tackled as a convex optimization problem~\cite{RebolloFD-TKDE10-tCloseToPRAMviaIT, CalmonF-Allerton12}.
However, in practice, model knowledge is often missing or inaccurate for realistic data sets, and the optimization becomes intractable for high-dimensional and continuous data.
Addressing these challenges, we propose a data-driven approach that optimizes the privacy-preserving mechanism to attain the theoretically optimal privacy-utility tradeoffs, by learning from a set of training data rather than requiring model knowledge.
We call this approach \textit{Privacy-Preserving Adversarial Networks} (PPAN)
since the mechanism, realized as a randomized neural network, is trained along with an adversarial network that attempts to recover the sensitive information from the released data.
The key to attaining information-theoretic privacy is that the adversarial network specifically estimates the posterior distribution (rather than only the value) of the sensitive variable given the released data to enable a variational approximation of mutual information~\cite{BarberA_IM-algo}.
While the adversary is trained to minimize the log-loss with respect to this posterior estimate,
the mechanism network is trained to attain the dual objectives of minimizing distortion and concealing sensitive information (by maximizing the adversarial loss).

\subsection{Related Work}

The general concept of adversarial training of neural networks was introduced by~\cite{goodfellow2014_GAN}, which proposed \textit{Generative Adversarial Networks} (GAN) for learning generative models that can synthesize new data samples.
Since their introduction, GANs have inspired a large and growing 
number of adversarially trained neural network architectures for a wide variety of purposes~\cite{GANZoo}.

The earlier works of~\cite{edwardsS15_ALFR, hamm_minimax},~\edits{\cite{hamm2017minimax}} have also proposed adversarial training frameworks for optimizing privacy-preserving mechanisms, where the adversarial network is realized as a classifier that attempts to recover a discrete sensitive variable.
In~\cite{edwardsS15_ALFR}, the mechanism is realized as an autoencoder, and the adversary attempts to predict a binary sensitive variable from the latent representation.
In the framework of~\cite{hamm_minimax},~\edits{\cite{hamm2017minimax}}, a deterministic mechanism is trained with the adversarial network realized as a classifier attempting to predict the sensitive variable from the output of the mechanism.
Both of these frameworks additionally propose using an optional predictor network that attempts to predict a useful variable from the output of the mechanism network.
Thus, while the adversarial network is trained to recover the sensitive variable, the mechanism and predictor (if present) networks are trained to realize multiple objectives: maximizing the loss of the adversary as well as minimizing the reconstruction loss of the mechanism network and/or the prediction loss of the predictor network.
However, a significant limitation of both of these approaches is that they consider only deterministic\footnote{While~\cite{hamm_minimax},~\edits{\cite{hamm2017minimax}} does also consider a ``noisy'' version of their mechanism, the randomization is limited to only independent, additive noise before or after deterministic filtering.} mechanisms, which generally do not achieve the optimal privacy-utility tradeoffs, although neither attempts to address information-theoretic privacy.
\edits{The work of~\cite{mirjalili2018semi} employs an adversarial framework similar to~\cite{edwardsS15_ALFR} to preserve gender-privacy of face images while retaining biometric recognition utility.
Within the broader context of empirical privacy measures addressed via adversarial training,~\cite{nasr2018machine} considers an adversarial framework for learning accurate predictive models that preserve the membership privacy of individuals that may be in the training dataset.}

The recent, independent work of~\cite{huang2017GAP} proposes a similar adversarial training framework, which also realizes the necessity of and proposes randomized mechanism networks, in order to address the information-theoretically optimal privacy-utility tradeoffs.
They also rediscover the earlier realization of~\cite{CalmonF-Allerton12} that mutual information privacy arises from an adversary (which outputs a distribution) that is optimized with respect to log-loss.
However, their framework does not make the connections to a general variational approximation of mutual information applicable to arbitrary (i.e., discrete, continuous, and/or multivariate) sensitive variable alphabets,
and hence their data-driven formulation and empirical evaluation is limited to only binary sensitive variables.

\subsection{Contributions and Paper Outline}

Our framework, presented in Section~\ref{sec:setup}, provides the first data-driven approach for optimizing privacy-preserving data release mechanisms that approaches the information-theoretically optimal privacy-utility tradeoffs.
A key novelty of our approach is the use of adversarial training to perform a variational approximation of mutual information privacy.
Unlike previous work, our approach can handle 
randomized data release mechanisms where the input to the mechanism can be a general observation of the data, e.g., a full or potentially noisy/partial view of the sensitive and useful variables.

In our proposed framework all of the variables that are involved can be discrete, continuous, and/or high-dimensional vectors. We develop specific network architectures and sampling methods appropriate for various scenarios in Section~\ref{sec:sampling}. In particular,
when all of the variables have finite alphabets, we demonstrate that the network architectures can be efficiently minimalized to essentially just the matrices describing the conditional distributions, and that replacing sampling with a directly computed expectation improves training performance.

We evaluate our PPAN approach in Section~\ref{sec:results} with experiments on 
synthetic data and the MNIST handwritten digit dataset. For the synthetic data experiment, we demonstrate that PPAN closely approaches the theoretically optimal privacy-utility tradeoff.
In Section~\ref{sec:discrete_exps}, we consider synthetic discrete-valued data following a \textit{symmetric pair} distribution and compare the privacy-utility tradeoff results with an approach addressing the same problem in \cite{MakhdoumiF-Allerton13}. 
In Section~\ref{sec:G_scalar}, we consider scalar jointly Gaussian sensitive and useful attributes and benchmark the performance of PPAN against the theoretically optimal privacy-utility tradeoff. 
In Section~\ref{sec:G_RDvec}, we demonstrate how the PPAN framework can be used to generate rate-distortion curves studied in information theory, purely from samples. Finally in Section~\ref{sec:PU} and Appendix~\ref{sec:proofs}, we provide and derive analytical expressions for the optimal privacy-utility tradeoffs for Gaussian distributed data and mean square error distortion. 
In the rest of the appendices, we present some extensions of our framework and visualizations.

%% file: framework.tex
\section{Problem Formulation and PPAN Methods}
\label{sec:setup}

\subsection{Privacy-Utility Tradeoff Optimization}
\label{sec:formulation}

\begin{figure}[h]
\centering
\includegraphics[width=0.48\textwidth]{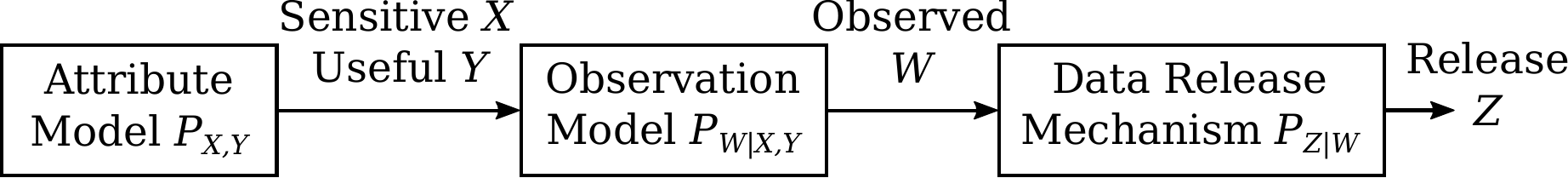}
\caption{Setting for privacy-utility tradeoff optimization.}
\label{fig:setup}
\end{figure}

We consider the privacy-utility tradeoff optimization problem
described in~\cite{ITA2016}, which extends the frameworks initiated
by~\cite{RebolloFD-TKDE10-tCloseToPRAMviaIT, CalmonF-Allerton12}. Figure~\ref{fig:setup} depicts the problem setting where 
observed data $W$, sensitive attributes $X$, and useful attributes $Y$
are modeled as random variables that are jointly distributed according
to a data model $P_{W,X,Y}$ over the space $\mathcal{W} \times
\mathcal{X} \times \mathcal{Y}$. The observed data $W$ is a potentially noisy/partial observation of the sensitive and useful data attributes $(X,Y)$.
The goal is to design and optimize the data release mechanism, i.e., a system that
processes the observed data $W$ to produce a release $Z \in
\mathcal{Z}$ that minimizes the privacy-leakage of the sensitive
attributes $X$, while also maximizing the utility gained from
revealing information about $Y$. This system is specified by the
\textit{release mechanism} $P_{Z|W}$, with $(W,X,Y,Z) \sim
P_{W,X,Y}P_{Z|W}$, and thus $(X,Y) \leftrightarrow W \leftrightarrow
Z$ forms a Markov chain.  Privacy-leakage is quantified by the mutual
information $I(X;Z)$ between the sensitive attributes $X$ and the
release $Z$.  Utility is inversely quantified by the expected
distortion $\mathbb{E}[d(Y,Z)]$
between the useful attributes $Y$ and the release $Z$, where the
distortion function $d: \mathcal{Y}\times \mathcal{Z} \to [0, \infty)$
is given by the application.  The design of the release mechanism
$P_{Z|W}$ is formulated as the following privacy-utility tradeoff
optimization problem,
\begin{equation}
\min_{P_{Z|W}: (X,Y) \leftrightarrow W \leftrightarrow Z} I(X;Z), \quad \text{s.t.} \quad \mathbb{E}[d(Y,Z)] \leq \delta, \label{eq:pp}
\end{equation} 
where the parameter $\delta$ indicates the distortion (or
\emph{disutility}) budget allowed for the sake of preserving privacy.

As noted in~\cite{ITA2016}, given a fixed data model $P_{W,X,Y}$ and
distortion function $d$, the problem in~\eqref{eq:pp} is a convex
optimization problem, since the mutual information objective $I(X;Z)$
is a convex functional of $P_{Z|X}$, which is in turn a linear
functional of $P_{Z|W}$, and the expected distortion
$\mathbb{E}[d(Y,Z)]$ is a linear functional of $P_{Y,Z}$ and hence
also of $P_{Z|W}$.  While the treatment in~\cite{ITA2016} considers
discrete variables over finite alphabets, the formulation
of~\eqref{eq:pp} need not be limited those assumptions.  Thus, in this
work, we seek to also address this problem with high-dimensional,
continuous variables. 
\edits{Although outside the focus of this work, in Appendix~\ref{sec:side_info} we discuss how mutual information privacy is impacted 
%the impact on mutual information privacy 
if there is some side information about $X$ available to an attacker.}
In Appendix~\ref{sec:MI-utility}, we discuss how to handle mutual information as a {\it utility} function within the PPAN framework as opposed to expected distortion that we focus on in this work.

\subsection{Adversarial Training for an Unknown Data Model}
\label{sec:CorePPAN}

\begin{figure}[h!]
\begin{center}
\includegraphics[width=0.48\textwidth]{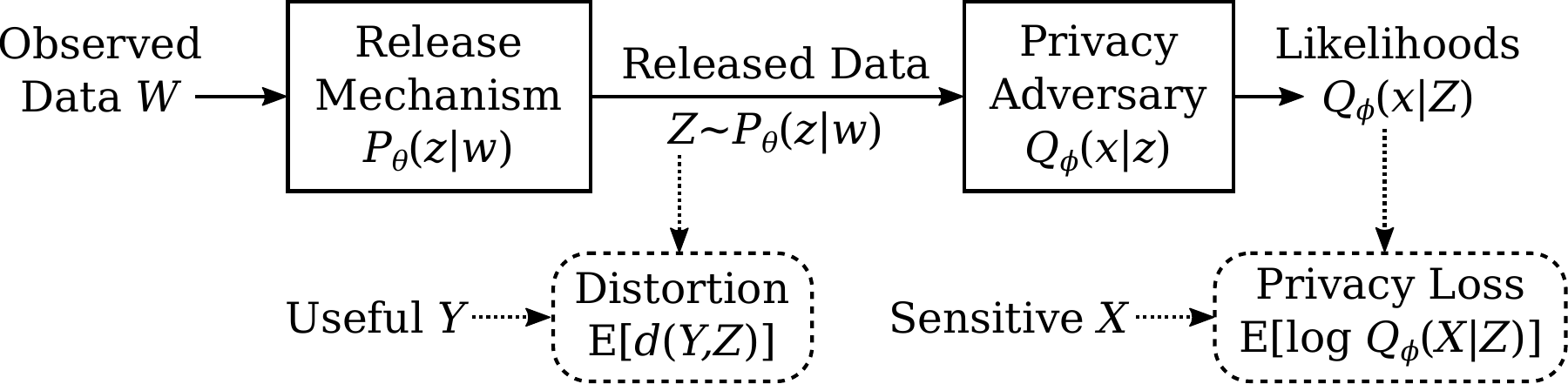}
\end{center}
\caption{Adversarial training framework.}
\label{fig:framework}
\end{figure}

Our aim is to solve the privacy-utility tradeoff optimization problem
when the data model $P_{W,X,Y}$ is unknown but instead a set of
training samples:
$\{(w_i, x_i, y_i)\}_{i=1}^n \sim \text{i.i.d.}\, P_{W,X,Y}$ \edits{is available}.\footnote{\edits{For the case when $X$ is not explicitly available during training, or it is vaguely defined, please see the discussion in Appendix~\ref{sec:surrogate}.}}  A key to our approach is approximating
$I(X;Z)$ via a variational lower bound given
by~\cite{BarberA_IM-algo} and also used in~\cite{ChenDHRSSA_infoGAN}.
This bound is based on the following identity which holds for any
distribution $Q_{X|Z}$ over $\mathcal{X}$ given values in
$\mathcal{Z}$
\begin{align*}
-h(X|Z) 
= \KL (P_{X|Z} \Vert Q_{X|Z}) + \mathbb{E} \big[\log Q_{X|Z}(X|Z)\big],
\end{align*}
where $\KL(\cdot\Vert\cdot)$ denotes the Kullback-Leibler (KL) divergence.  Therefore, since $I(X;Z) = h(X)-h(X|Z)$ and
KL divergence is nonnegative,
\begin{align}\label{eq:MI_var_lb}
h(X) + \max_{Q_{X|Z}} \mathbb{E} \big[\log Q_{X|Z}(X|Z) \big] =
I(X;Z),
\end{align}
where the maximum is attained when the variational posterior $Q_{X|Z}
= P_{X|Z}$.  Using~\eqref{eq:MI_var_lb} with the constant $h(X)$ term
dropped, we convert the formulation of~\eqref{eq:pp} to an
unconstrained minimax optimization problem,
\begin{align}\label{eq:min-max}
\min_{P_{Z|W}} \max_{Q_{X|Z}}\mathbb{E}\big[\log Q_{X|Z}(X|Z)\big] +
\lambda \mathbb{E}\big[d(Y,Z)\big],
\end{align}
where the expectations are with respect to $(W,X,Y,Z) \sim P_{W,X,Y}
P_{Z|W}$, and the parameter $\lambda > 0$ can be adjusted to obtain
various points on the optimal privacy-utility tradeoff curve.
Alternatively, to target a specific distortion budget $\delta$, the
second term in~\eqref{eq:min-max} could be replaced with a penalty
term $\lambda (\max(0, \mathbb{E}[d(Y,Z)] - \delta))^2$, where $\lambda
> 0$ is made relatively large to penalize exceeding the budget.  The
expectations in~\eqref{eq:min-max} can be conveniently approximated by Monte Carlo sampling over training set batches.

The minimax formulation of~\eqref{eq:min-max} can be interpreted and
realized in an adversarial training framework (as illustrated by
Figure~\ref{fig:framework}), where the variational posterior $Q_{X|Z}$
is viewed as the posterior likelihood estimates of the sensitive attributes $X$
made by an adversary observing the release $Z$.  
The data release mechanism is trained to minimize both the distortion and privacy loss terms, while the adversary is trained to maximize the privacy loss. 
Specifically, the adversary attempts to maximize the negative log-loss $\mathbb{E}[\log Q_{X|Z}(X|Z)]$, which the release mechanism $P_{Z|W}$ attempts to minimize.  The release mechanism and adversary are realized as neural
networks, which take as inputs $W$ and $Z$, respectively, and produce
the parameters that specify their respective distributions $P_{Z|W}$
and $Q_{X|Z}$ within parametric families that are appropriate for the
given application.  For e.g., a release mechanism suitable for the
release space $\mathcal{Z} = \mathbb{R}^d$ could be the multivariate
Gaussian
\begin{equation*}
P_{Z|W}(z|w) = \mathcal{N}(z; (\boldsymbol{\mu}, \boldsymbol{\Sigma}) = f_\theta(w)),
\end{equation*}
where the mean $\boldsymbol{\mu}$ and covariance $\boldsymbol{\Sigma}$
are determined by a neural network $f_\theta$ as a function of $w$ and
controlled by the parameters $\theta$.  For brevity of notation, we
will use $P_\theta(z|w)$ to denote the distribution defined by the
release mechanism network $f_\theta$.  Similarly, we will let
$Q_\phi(x|z)$ denote the parametric distribution defined by the adversary
network that is controlled by the parameters $\phi$.  For each
training sample tuple $(w_i, x_i, y_i)$, we sample $k$ independent
releases $\{z_{i,j}\}_{j=1}^k \stackrel{\text{iid}}{\sim}
P_\theta(z|w_i)$ to approximate the loss term with
\begin{align}\label{eq:PPAN_loss}
\mathcal{L}^i(\theta, \phi) := \frac{1}{k} \sum_{j=1}^k \left[\log
  Q_\phi(x_i|z_{i,j}) + \lambda d(y_i,z_{i,j}) \right].
\end{align}
The networks are optimized with respect to these loss terms averaged
over the training data (or mini-batches)
\begin{align}\label{eq:PPANobj}
\min_\theta \max_\phi \frac{1}{n}\sum_{i=1}^n \mathcal{L}^i(\theta,
\phi),
\end{align}
which approximates the theoretical privacy-utility tradeoff
optimization problem as given in~\eqref{eq:min-max}, since by the law
of large numbers, as $n \rightarrow \infty$,
\begin{align*}
\frac{1}{n}\sum_{i=1}^n \mathcal{L}^i(\theta, \phi) 
&\xrightarrow{\mathrm{a.s.}} \mathbb{E}\big[\log Q_\phi(X|Z) + \lambda
  d(Y,Z)\big],
\end{align*}
where the expectation is with respect to $(W,X,Y,Z) \sim
P_{W,X,Y}P_\theta(z|w)$.  Similarly, the second term
in~\eqref{eq:PPAN_loss} could be replaced with a penalty term $\lambda
(\max(0, d(y_i,z_{i,j}) - \delta))^2$ to target a specific distortion
budget $\delta$.  Similar to GANs~\cite{goodfellow2014_GAN}, the
minimax optimization in~\eqref{eq:PPANobj} can be more practically
handled by alternating gradient descent/ascent between the two
networks (possibly with multiple inner maximization updates per outer
minimization update) rather than optimizing the adversary network
until convergence for each release mechanism network update.
See Appendix~\ref{sec:pseudocode} 
for the pseudocode description of the algorithm.

\subsection{Sampling the Release Mechanism}
\label{sec:sampling}

To allow optimization of the networks via gradient methods, the
release samples need to be generated such that the gradients of the
loss terms can be readily calculated.  Various forms of the release
mechanism distribution $P_\theta(z|w)$ are appropriate for different
applications, and each require their own specific sampling methods. \edits{Finite alphabet models are appropriate for categorical data such as star ratings and quantized census data whereas Gaussian, mixture of Gaussian or more general real-valued models are more appropriate for voice, image, video, and other physical sensor data.}

\subsubsection{Finite Alphabets} \label{sec:finite_arch}

When the release space $\mathcal{Z}$ is a finite discrete set, we can
forgo sampling altogether and calculate the loss terms via
\begin{align}\label{eqn:finite_arch}
\mathcal{L}^i_{\text{disc}}(\theta, \phi)
:= \sum_{z\in\mathcal{Z}}
P_\theta(z|w_i) (\log Q_\phi(x_i|z) + \lambda d(y_i,z)),
\end{align}
which replaces the empirical average over $k$ samples with the direct
expectation over $Z$. We found that this direct expectation produced
better results than estimation via sampling, such as by applying the
Gumbel-softmax categorical reparameterization trick
(see~\cite{maddison2016concrete, jang2016categorical}).
\edits{Here we assume that the alphabet size is known. % and the training samples are
Since this is a data-driven mechanism, we will obtain good performance if the empirical distribution of the training data does not diverge much from the actual unknown dataset distribution.
This is often a standard assumption in different setups, for e.g., in the Probably Approximately Correct (PAC) notion of learning.}
\edits{In practice, Bayesian priors for the estimation of the conditional distributions that appear in~\eqref{eqn:finite_arch} could also be incorporated in order to mitigate the curse-of-dimensionality issue wherein the alphabet sizes are much larger than the size of the training set.}

Further, if $\mathcal{W}$ and $\mathcal{X}$ are also finite alphabets,
then $P_\theta(z|w)$ and $Q_\phi(x|z)$ can be exactly parameterized by
matrices of size $|\mathcal{Z}| \times |\mathcal{W}|$ and
$|\mathcal{X}| \times |\mathcal{Z}|$, respectively.  Thus, in the
purely finite alphabet case, with the variables represented as one-hot
vectors, the mechanism and adversary are most efficiently realized as
networks with no hidden layers and softmax applied to the
output (to yield stochastic vectors).

\subsubsection{Gaussian Approximations for Reals} 
\label{subsec:gaussian_release}

A multivariate Gaussian release mechanism can be sampled by employing
the reparameterization trick of~\cite{kingma2013-VAE}, which first
samples a vector of independent standard normal variables $\mathbf{u}
\sim \mathcal{N}(\mathbf{0}, \mathbf{I})$, and then generates $z =
\mathbf{A}\mathbf{u} + \boldsymbol{\mu}$, where the parameters
$(\boldsymbol{\mu}, \mathbf{A}) = f_\theta(w)$ are produced by the
release mechanism network to specify a conditional Gaussian with mean
$\boldsymbol{\mu}$ and covariance $\boldsymbol{\Sigma} =
\mathbf{A}\mathbf{A}^T$. 
This approach can be extended to Gaussian Mixture Models as explained in  Appendix~\ref{sec:GMM}.
\subsubsection{Universal Approximators} \label{sec:universal_approx}

Another approach, as seen in~\cite{makhzani2015adversarial}, is to
directly produce the release sample as $z = f_\theta(w, u)$ using a
neural network that takes random seed noise $u$ as an additional
input.  The seed noise $u$ can be sampled from a simple distribution
(e.g., uniform, Gaussian, etc.) and provides the randomization of $z$
with respect to $w$.  Since the transformations applying the seed
noise can, in principle, be learned, this approach could potentially approximate any ``nice'' distribution due to the universal approximation properties of neural networks.
However, although it is not needed
for training, it is generally intractable to produce an explicit
expression for $P_\theta(z|w)$ as implied by the 
network.

%% file: results.tex
\section{Experimental Results}\label{sec:results}
\input{table}
In this section, we present the privacy-utility tradeoffs that are achieved by our PPAN framework in experiments with synthetic and real data.
For the synthetic data experiments, we show that the results obtained by PPAN (which does not require model knowledge and instead uses training data) are very close to the theoretically optimal tradeoffs obtained from optimizing~\eqref{eq:pp} with full model knowledge.
In the experiments with discrete synthetic data presented in Section~\ref{sec:discrete_exps}, we also compare PPAN against the approach of~\cite{MakhdoumiF-Allerton13}, where first an approximate discrete distribution is estimated from the training data, which is then used in place of the true distribution for the optimization in~\eqref{eq:pp}.
This two-step procedure involves model estimation as its first step, and is in general not tractable for high-dimensional continuous distributions. 
For the synthetic data experiment, we consider Gaussian joint distribution over the sensitive, useful, and observed data, for which we can compare the results obtained by PPAN against the theoretically optimal tradeoffs (derived in Section~\ref{sec:PU}).
We use the MNIST handwritten digits dataset to illustrate the application of the PPAN framework 
to real data in Section~\ref{sec:mnist}.
We demonstrate optimized networks that can trace the tradeoff between concealing the digit and reducing image distortion.
Table~\ref{tab:models} summarizes the data models and distortion metrics that we use in our experiments.
Our experiments were implemented using the Chainer deep learning framework~\cite{chainer}, with optimization performed by their implementation of Adam~\cite{kingma2014adam}.
\edits{We used the Chainer-default Adam parameters in all of our experiments: $\alpha=0.001$, $\beta_1=0.9$, $\beta_2=0.999$, $\epsilon=10^{-8}$.}

\subsection{Discrete Synthetic Data} \label{sec:discrete_exps}
\input{discrete-results}

%% file: table.tex
\begin{table*}[t!]
\footnotesize
\begin{center}
\caption{The models used for obtaining synthetic training and test datasets
  in our experiments.}
\label{tab:models}
{\renewcommand{\arraystretch}{1.2}
\begin{tabular}{c|P{4.5cm}|P{2cm}|c}
\hline
Case & Attribute Model & Observation & Distortion Metric
\\ \hline \hline
Discrete, Sec.~\ref{sec:discrete_exps} & $(X,Y)$ symmetric pair for $m=10, p=0.4$, see
\eqref{eq:symmetric_pmf1} & $W = Y$ and $W = (X,Y)$ & $\Pr[Y \neq Z]$
\\ \hline
Continuous, Sec.~\ref{sec:G_OPvec} %sec:continuous_exps
& $\begin{bsmallmatrix} X\\ Y \end{bsmallmatrix} \sim
\mathcal{N}\left(\bm{0}, \begin{bsmallmatrix} I_5 &
  \diag(\bm{\rho})\\ \diag(\bm{\rho}) & I_5\end{bsmallmatrix}
  \right)$, $\bm{\rho} = [0.47, 0.24, 0.85, 0.07, 0.66]$ & $W = Y$
  & $\mathbb{E}[\| Y - Z \|^2]$ \\ \hline
Continuous, Sec.~\ref{sec:G_scalar} & $\begin{bsmallmatrix} X\\ Y \end{bsmallmatrix} \sim
\mathcal{N}\left(\bm{0}, \begin{bsmallmatrix} 1 & 0.85\\ 0.85 &
  1\end{bsmallmatrix} \right)$
  & $W = Y$ and $W = (X,Y)$ & $\mathbb{E}[(Y-Z)^2]$\\ \hline
Continuous, Sec.~\ref{sec:G_RDvec} & $X = Y \sim\mathcal{N}\left(\bm{0}, \diag(\bm{\sigma}^2) \right)$,
 $\bm{\sigma}^2 = [0.47, 0.24, 0.85, 0.07, 0.66]$
& $W = X = Y$ & $\mathbb{E}[\| Y - Z \|^2]$\\ \hline
\end{tabular}}
\end{center}
\vspace{-2.5ex}
\end{table*}

%% file: discrete-results.tex
\begin{figure}
\centering
\includegraphics[scale=0.47]{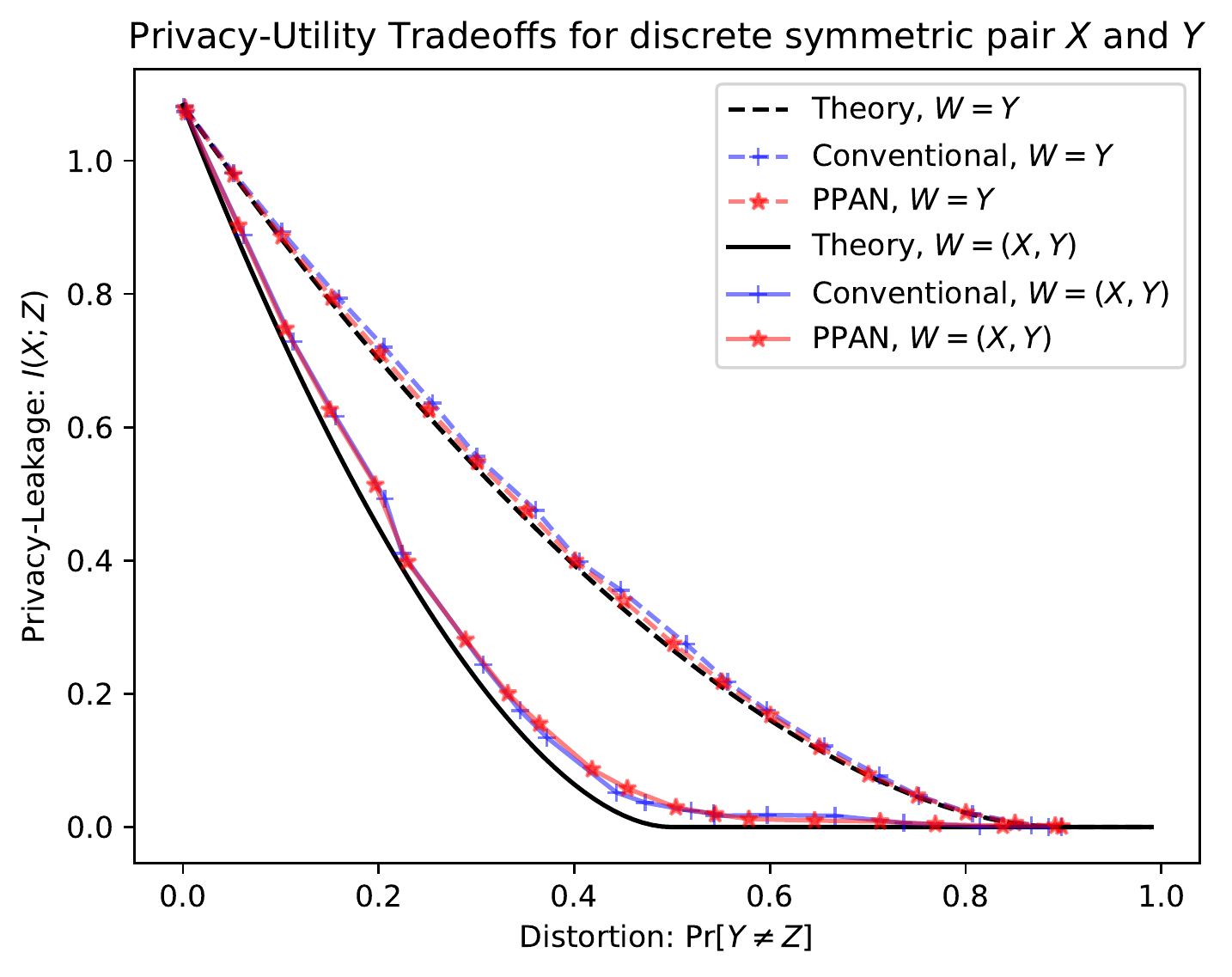}
\vglue -2 ex
\caption{Comparison of PPAN performance against the conventional model estimation approach of~\cite{MakhdoumiF-Allerton13} and the theoretical optimum,
for two observation scenarios: full data observed, i.e. $W=(X,Y)$, shown by solid lines, and only useful attribute observed, i.e. $W=Y$, shown by dashed lines.
\label{fig:discrete}}
\end{figure}

In our experiments with discrete data, we will consider two observation models, full data (where $W=(X,Y)$) and useful data only (where $W=Y$). We use a toy distribution for the attributes for which the theoretically optimal privacy-utility tradeoffs have been analytically derived in~\cite{wang2017privacy}, 
using probability of error as the distortion metric, i.e.,  $\mathbb{E}[\mathbf{1}(Y \neq Z)] = \Pr[Y \neq Z]$.
Specifically, we consider sensitive and useful attributes that are distributed over the finite alphabets $\mathcal X = \mathcal Y = \{0,\ldots,m-1\}$, with $m {\geq} 2$ and parameter $p \in [0,1]$, according to the {\em symmetric pair} distribution given by
\begin{align}\label{eq:symmetric_pmf1}
P_{X,Y}(x,y) =
\begin{cases}
\frac{1-p}{m}, & \text{if } x=y, \\
\frac{p}{m (m-1)}, & \text{otherwise}.
\end{cases}
\end{align}
\subsubsection{Network Architecture and Evaluation}

As mentioned in Section~\ref{sec:finite_arch}, 
the network architecture for the release mechanism and adversary can be reduced to a bare minimum when all of the variables are finite-alphabet.
Each network simply applies a single linear transformation (with no bias term) on the one-hot encoded input, followed by the softmax operation to yield a stochastic vector.
The mechanism network takes as input $w$ encoded as a one-hot column vector $\mathbf{w}$ and outputs
$P_\theta(\cdot | w) = \mathrm{softmax}(\mathbf{Gw})$,
where the network parameters $\theta = \mathbf{G}$ are the entries of a $|\mathcal{Z}| \times |\mathcal{W}|$ real matrix.
Note that applying the softmax operation to each column of $\mathbf{G}$ produces the conditional distribution $P_{Z|W}$ describing the mechanism.
Similarly, the attacker network is realized as
$Q_\phi(\cdot | z) = \mathrm{softmax}(\mathbf{Az})$,
where $\mathbf{z}$ is the one-hot encoding of $z$, and the network parameters $\phi = \mathbf{A}$ are entries of a $|\mathcal{X}| \times |\mathcal{Z}|$ real matrix.
We optimize these networks according to~\eqref{eq:PPANobj}, using the penalty term modification of the loss terms in~\eqref{eqn:finite_arch} as given by
\begin{align*}
&\mathcal{L}^i_{\text{disc}}(\theta, \phi) :=\\
&\sum_{z\in\mathcal{Z}} P_\theta(z|w_i) \big( \log Q_\phi(x_i|z) + \lambda \max(0, d(y_i,z) - \delta)^2 \big).
\end{align*}
We use $\lambda = 500$ in these experiments.

In Figure~\ref{fig:discrete}, we compare the results of PPAN against the theoretical baselines 
given by~\cite{wang2017privacy} (\textit{c.f.} Appendix~\ref{sec:symm-pair-results}), as well as against a conventional approach suggested by~\cite{MakhdoumiF-Allerton13}, where the joint distribution of $P_{W,X,Y}$ is estimated from the training data and then used in the convex optimization of~\eqref{eq:pp}. We can see that the PPAN mechanism learns a data release distribution 
that has close to optimal privacy leakage for a wide range of distortion values. 
We used $1000$ training samples generated according to the symmetric pair distribution in~\eqref{eq:symmetric_pmf1} with $m = 10$ and $p = 0.4$.
The PPAN networks were trained for $2500$ epochs (for the full data observation case) with a minibatch size of $100$, with each network alternatingly updated once per iteration.
For the useful data only observation case, $2000$ epochs were used.
For evaluating both the PPAN and conventional approaches, we computed the mutual information and probability of error from the joint distribution that combines the optimized $P_{Z|W}$ with the true $P_{X,Y,W}$.

%% file: cont-results.tex
\subsection{Gaussian Synthetic Data} \label{sec:continuous_exps}

In this section, we consider scalar and multivariate jointly Gaussian sensitive and useful
attributes. We evaluate the performance of PPAN on synthetic data
following this model in various scenarios. The distortion metric
is the mean squared error between the release and the useful
attribute.
As we note in Section~\ref{sec:PU}, the optimum release 
for the scenarios considered here 
is jointly Gaussian with the attributes. Thus we could use a
mechanism network architecture that can realize the procedure
described in Section~\ref{subsec:gaussian_release} to generate the
release. However, since the optimal release distribution is not known
for general attribute models, we use the universal approximator technique described in
Section~\ref{sec:universal_approx}.

The mechanism implemented in these experiments consists of
three fully connected layers, with the ReLU activation function
applied at the outputs of the two hidden layers, and no activation function is
used at the output layer. The mechanism takes as input
observation $W$ and seed noise $U$, and generates the
release $Z = f_\theta(W, U)$, where
$\theta$ denotes the parameters of the mechanism network.
Components of the seed noise vector are i.i.d. Uniform$[-1,1]$. 
The adversary network, with parameters denoted by $\phi$,
models the posterior probability $Q_\phi(X|Z)$
of the sensitive attribute given the release. We assume that
$Q_\phi(\cdot |z)$ is a normal distribution with mean vector 
$\bm{\mu}_\phi(z)$ and covariance matrix $\diag(\bm{\sigma}^2_\phi(z))$, i.e.,
they are functions of the release $z$.
The adversary network has three fully connected layers to learn
the mean and variances. The network takes as input the release $z$ and
outputs the pair $(\bm{\mu}_\phi(z), \log
\bm{\sigma}^2_\phi(z))$, where the $\log$ is applied componentwise on
the variance vector. 
We use the adversarial networks to solve the min-max optimization problem 
described in~\eqref{eq:PPANobj}. We choose $k=1$ in~\eqref{eq:PPAN_loss}, 
and similar to the previous section, we use the penalty modification of 
the distortion term, i.e., 
\begin{equation} \label{eqn:gauss-loss}
\mathcal{L}^i_{\text{gauss}}(\theta, \phi) = \log Q_{\phi}(x_i|z_i) + 
\lambda (\max(0, \lVert y_i - z_i\rVert^2 - \delta))^2.
\end{equation}
The parameter $\delta$ is 
swept through a linearly spaced range of values. 
The values chosen for the multiplier $\lambda$ and the distortion budget $\delta$ in various experiments is described in the sections below. 
For each value of $\delta$, we sample the data model to obtain an 
independent dataset realization and use it to train and test the 
adversarial networks. We use $8000$ training samples and evaluate 
the performance of PPAN on $4000$ test samples.
For the scalar data experiments, both networks have 5 nodes per hidden layer,
while 20 nodes per hidden layer were used for the multivariate data experiments.
Each hidden layer has 20 nodes. The adversarial networks were trained for 250 epochs with a minibatch size of 200. In each
iteration we do 5 gradient descent steps to update the parameters of
the adversary network before updating the mechanism network. 

\subsubsection{Estimating Mutual Information Leakage}\label{subsec:MI_estimate}
\input{mutual-info-estimate}

\subsubsection{Rate Distortion} \label{sec:G_RDvec}
\input{rate_distortion}

\subsubsection{Scalar Gaussian Attributes} \label{sec:G_scalar}
\input{scalar-gaussian-results}

\subsubsection{Multivariate Gaussian Attributes} \label{sec:G_OPvec}

\begin{figure}
\centering
    \includegraphics[scale=0.45]{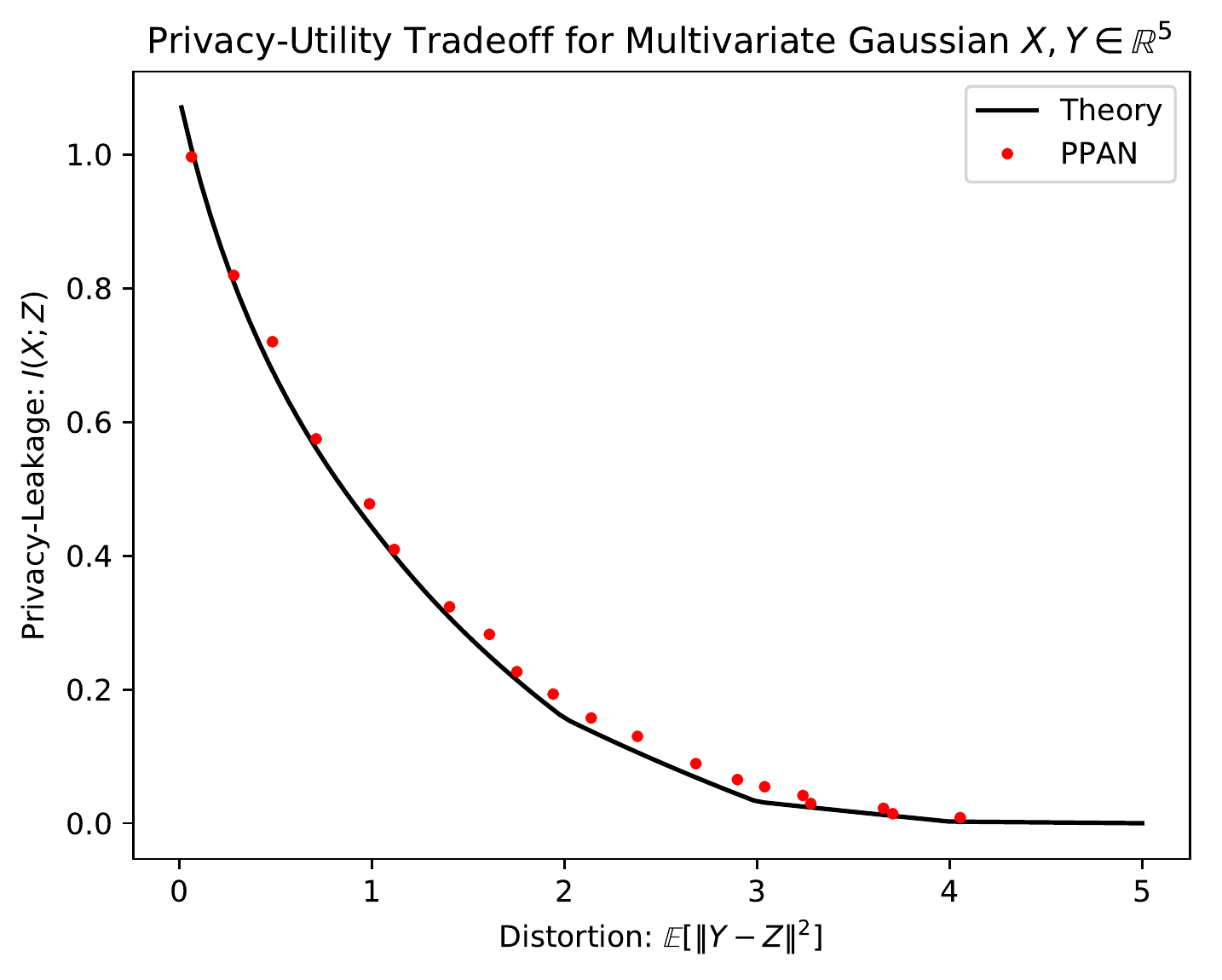}
    \caption{Comparison of the results achieved by PPAN versus the theoretical optimum tradeoff curve for the useful data only observation model (i.e., $W=Y$) for multivariate Gaussian $(X,Y)$.}
    \label{fig:vector_OP}
\end{figure}

Consider multivariate jointly Gaussian sensitive and useful attributes
$\begin{bsmallmatrix} X\\ Y \end{bsmallmatrix} \sim
\mathcal{N}\left(\bm{0}, \begin{bsmallmatrix} I_5 &
\diag(\bm{\rho})\\ \diag(\bm{\rho}) & I_5\end{bsmallmatrix} \right)$
where both $X, Y \in \mathbb{R}^5$ and $\bm{\rho} = [0.47, 0.24, 0.85, 0.07, 0.66]$. The observation model is UD, i.e., $W = Y$. 
We choose the multiplier $\lambda = 10$ and $20$ linearly spaced values for $\delta$ in the range $[0, 4.5]$. 
The seed noise is a vector with $8$ components. 
We plot the privacy-leakage and distortion values returned by the PPAN
mechanism on the test set along with the optimal tradeoff curve (from
Proposition~\ref{propn:vector_OP}) in Figure~\ref{fig:vector_OP}. The privacy-leakage values were estimated following the procedure in Section~\ref{subsec:MI_estimate}. 
The performance of the PPAN mechanism is very close to the 
theoretically optimum tradeoff curve over a wide range of target distortion
values. We visualize the true data attributes and the released attributes obtained by a trained PPAN mechanism in Figure~\ref{fig:scatter_viz} of Appendix~\ref{sec:multivar_Gdata_viz}. 

%% file: mutual-info-estimate.tex
Distortion caused by a release is estimated by the empirical 
mean squared error with respect to the testing samples.
However, estimating mutual information to evaluate privacy leakage is less straightforward
since the joint distribution $P_{X,Z}$ as realized by the optimized mechanism 
is not available explicitly. 
Since for these experiments, the optimal release $Z$ is jointly
Gaussian with $X$ (as we show in Section~\ref{sec:PU}), we estimate
$I(X; Z)$ via a Gaussian approximation.
Specifically, we use the expression for the mutual information of jointly Gaussian random vectors and replace all covariance matrices that appear there by their empirical counterparts, i.e.,
$\hat{I}(X; Z) = 
0.5 \log(\det (\hat{\Sigma}_X)/
\det (\hat{\Sigma}_{X | Z}))$,
where 
$\hat{\Sigma}_{X | Z} :=
\hat{\Sigma}_X - \hat{\Sigma}_{X, Z} \hat{\Sigma}_Z^{+}
\hat{\Sigma}_{X, Z}^T$
and $\hat{\Sigma}_X$ denotes the empirical self covariance matrix of $X$, $\hat{\Sigma}_Z^+$ denotes the pseudoinverse of the empirical self covariance matrix of $Z$, and $\hat{\Sigma}_{X, Z}$ denotes their empirical cross covariance matrix.
This underestimates the true
mutual information leakage since
\begin{align*}
I(X; Z) 
%&= h(X) -h(X|Z) \\
&= h(X) - h(X - \widehat{\mathbb{E}}[X|Z]|Z)\\
&\geq h(X) - h(X - \widehat{\mathbb{E}}[X|Z]) = \hat{I}(X;Z),
\end{align*}
where $\widehat{\mathbb{E}}[X|Z]$ is the
linear MMSE estimate of $X$ as a function of
$Z$. We use this estimate only for its simplicity, and one
could use other non-parametric estimates of mutual
information \cite{pmlr-v22-poczos12}.

%% file: rate_distortion.tex
We can apply the PPAN framework to the problem of computing
the minimum required rate of a code that describes a multivariate source $X$ to within a target value of expected distortion.  
This is a standard problem
in information theory when the source distribution is known, for example,
see Chapter~10 of~\cite{cover2012elements}.
However, the PPAN framework can be used to empirically approximate the rate-distortion curve from i.i.d. samples of the source without knowledge of the source distribution. The computation of the rate-distortion function can be viewed as
a degenerate case of the PPAN framework with $W = X = Y$, i.e., the
sensitive and useful attributes are the same and the observed dataset
is the attribute. The release $Z$ corresponds to an estimate $\hat{X}$
with expected distortion
less than a target
level while retaining as much expected uncertainty about $X$ as possible.

We illustrate the PPAN approach using a Gaussian source $X \in \mathbb{R}^5$ and mean squared error distortion.
For the experiment, we choose the attribute model $X \sim \mathcal{N}(\bm{0},
\diag(0.47, 0.24, 0.85, 0.07, 0.66))$ and the value $\lambda =500$. 
We run the experiment for $20$ different values of the target
distortion, linearly spaced between 0 to 2.5. The inputs to the
adversarial network are realizations of the attributes and seed noise. 
The seed noise is chosen to be a random vector of length 8 with each 
component i.i.d. Uniform$[-1,1]$. The network architecture 
and values of other hyperparameters are the same as those used for 
multivariate Gaussian attributes in Section~\ref{sec:continuous_exps}. 
Using the learned parameters $\theta^\star$, 
the mechanism network generates a release as
$Z=f_{\theta^\star}(W, U)=f_{\theta^\star}(X, U)$.
The distortion is estimated by the empirical mean squared error of the release 
with respect to the training samples. The privacy loss is quantified
by the estimate $\hat{I}(X; Z)$ as described in
Section~\ref{subsec:MI_estimate}.

The optimal privacy-utility tradeoff (or, rate-distortion) curve is
$R(D) = \min_{P(Z|X) ~:~ \mathbb{E}||X - Z||^2 \leq D} I(X;Z) = \sum_{j=1}^5 \max\{0, 0.5 \log ((\bm{\sigma}[j])^2 / D_j)\}$
\cite{cover2012elements}, where 
$\bm{\sigma}^2$ are the true variance parameters of the attribute distribution 
and $\sum_{j=1}^{5}D_j = D$. The values of $D_j$ for each component is
obtained using the Karush-Kuhn-Tucker (KKT) conditions for the constrained 
optimization problem, the solution of which is a standard waterfilling procedure. 

\begin{figure}
\centering
\includegraphics[scale=0.43]{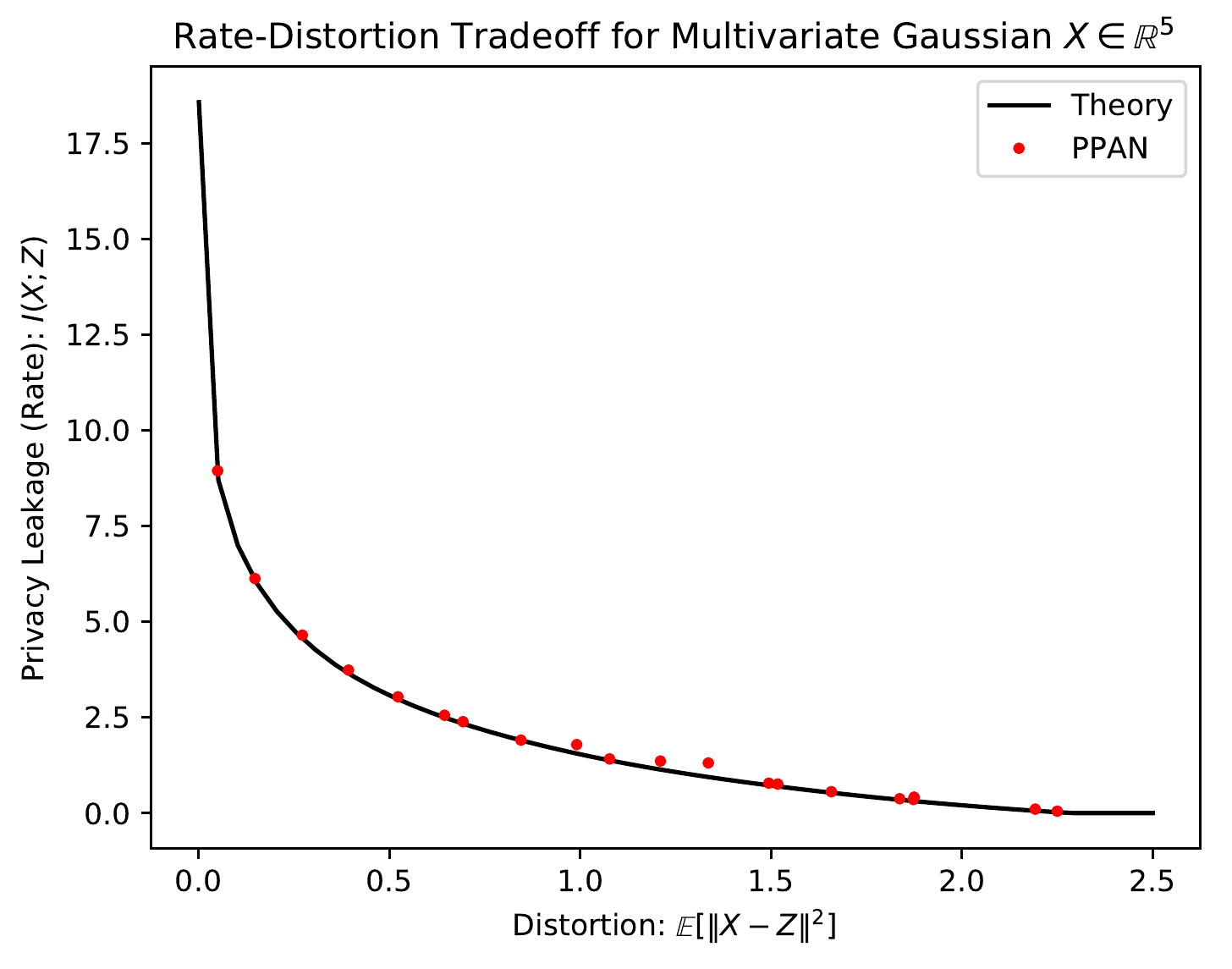}
\caption{Comparison of results obtained by PPAN versus the the optimal rate-distortion curve, for the rate-distortion problem where $X = Y = W$ is multivariate Gaussian.
}\label{fig:vector_RD}
\end{figure}

We plot the (privacy-leakage,
utility loss) pairs returned by the PPAN mechanism along with the
optimal tradeoff curve in Figure~\ref{fig:vector_RD}. One can see that
the operating points attained by the PPAN mechanism are very close to the theoretical
optimum tradeoff for a wide range of target distortion values.

%% file: scalar-gaussian-results.tex
\begin{figure}
\centering
\includegraphics[scale=0.45]{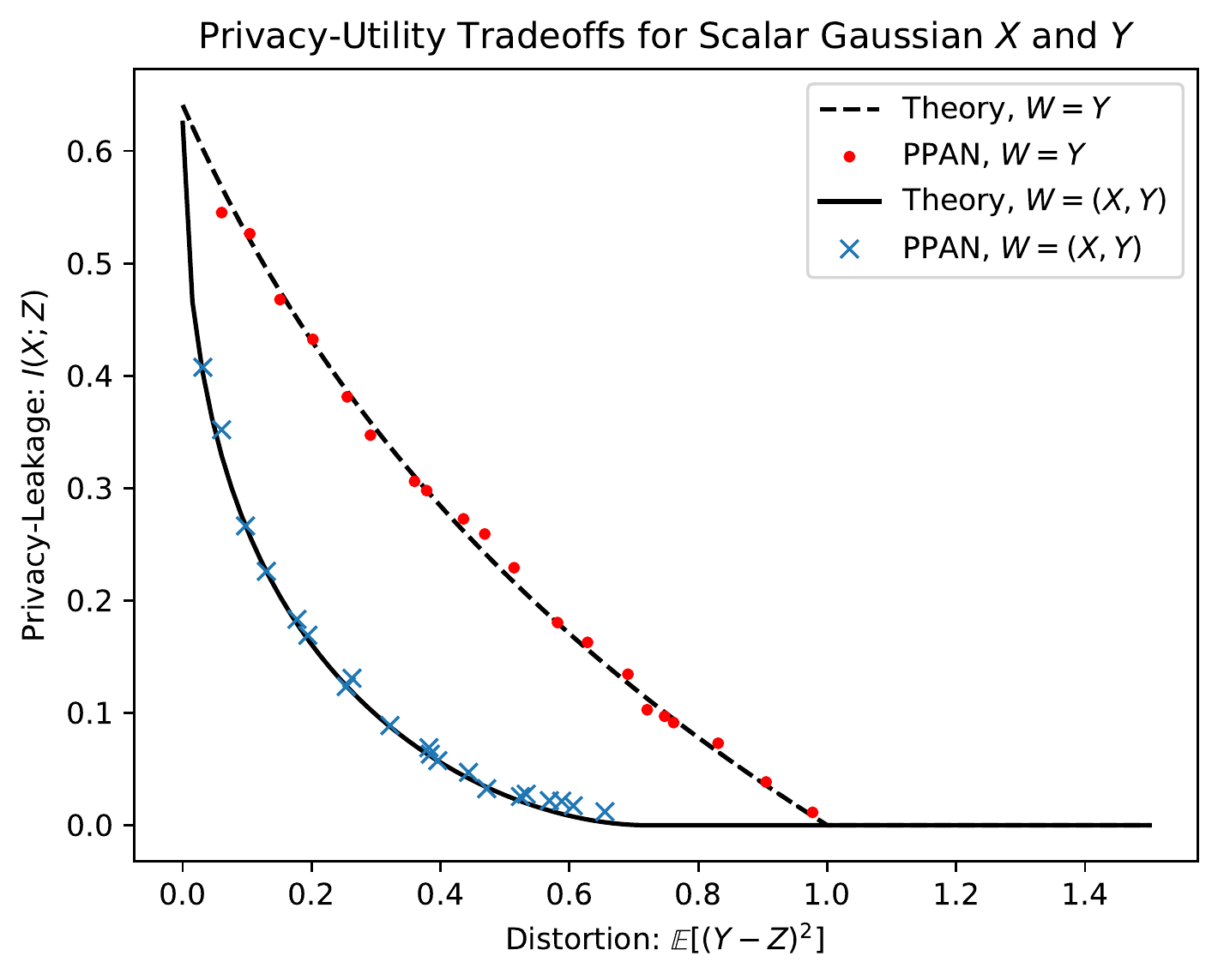}
\vglue -2ex
\caption{Comparison of the results achieved by PPAN versus the theoretical optimum tradeoff curve, with jointly Gaussian scalar $(X,Y)$, for the useful data only (i.e., $W=Y$) and the full data (i.e., $W=(X,Y)$) observation models.
}\label{fig:scalar_FD_OP}
\vglue -1ex
\end{figure}

Consider jointly Gaussian sensitive and useful attributes such that 
$\begin{bsmallmatrix} X \\ Y\end{bsmallmatrix} \sim
  \mathcal{N}\big(\begin{bsmallmatrix}0
    \\ 0 \end{bsmallmatrix}, \begin{bsmallmatrix}1 & 0.85\\ 0.85 &
    1\end{bsmallmatrix} \big)$.
We analyze two different observation models here: $W=Y$, called useful data only (UD)
and $W=(X,Y)$, called full data (FD). The distortion metric is the mean squared error between the release and the useful attribute.
The seed noise is a scalar random variable following Uniform$[-1,1]$. 
The values of the multipliers chosen are:
$\lambda^{\text{UD}} = 10$ and $\lambda^{\text{FD}} = 50$.    
In each case, we run experiments for 20 different values of the target distortion
with $\delta^{\text{UD}} \in [0,1]$ and $\delta^{\text{FD}} \in
[0,0.8]$.
The privacy-leakage and distortion 
values returned by the PPAN mechanism on the test set are plotted along 
with the optimal tradeoff curves (from Propositions~\ref{propn:scalar_OP} and
\ref{propn:scalar_FD}) in Figure~\ref{fig:scalar_FD_OP}. In both the
observation models, we observe that the PPAN mechanism generates
releases that have nearly optimal privacy-leakage over a range of
distortion values.

%% file: mnist.tex
\subsection{MNIST Handwritten Digits} \label{sec:mnist}

\begin{figure}
\centering
\includegraphics[width=0.45\textwidth]{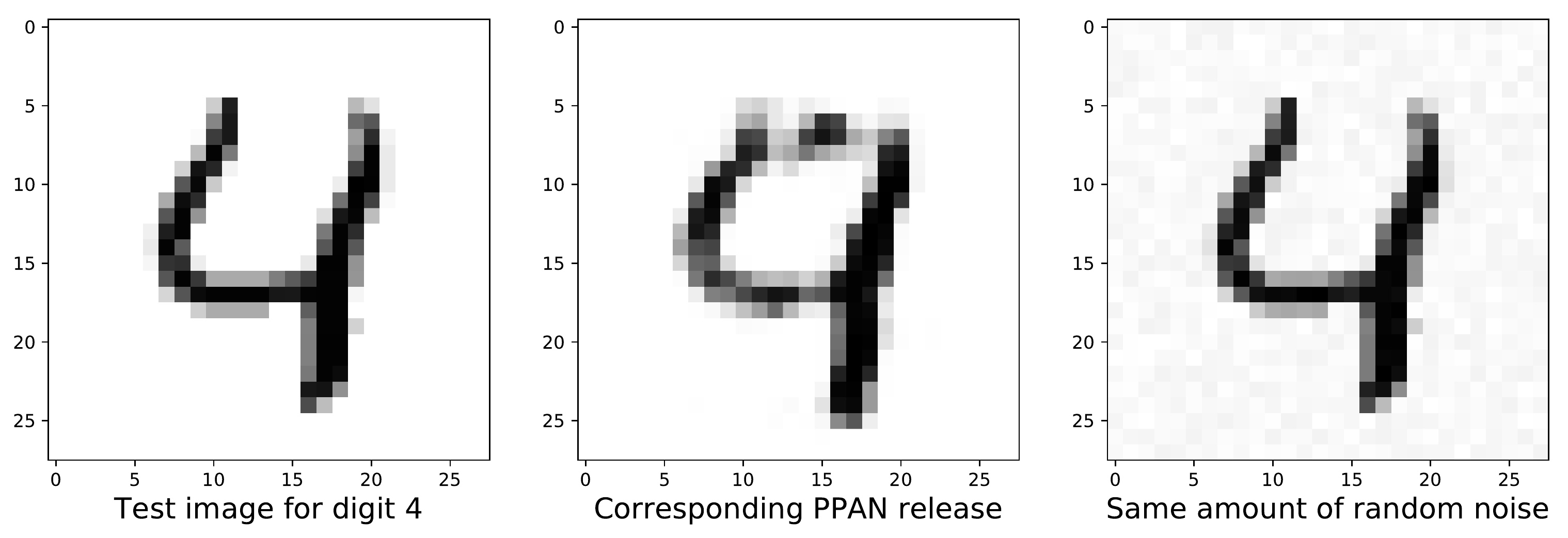}
\caption{A digit `$4$' from the MNIST test set and the release generated by a PPAN mechanism trained at $\lambda=25$. Adding random noise to each pixel for the same amount of total noise results in the third image.}
\label{fig:mnist_4}
\end{figure}

The MNIST dataset consists of 70K labeled images of handwritten digits split into training and test sets of 60K and 10K images, respectively.
Each image consists of $28 \times 28$ grayscale pixels, which we handle as vectors in $[0,1]^{784}$.
In the first set of experiments,
we consider the image to be both the useful and the observed data, i.e., $W = Y$, the digit label to be the sensitive attribute $X$, and the mechanism release as an image $Z \in [0,1]^{784}$.
We measure the 
distortion between the original and released images $Y, Z$ as 
\begin{align*}
d(Y, Z) := \frac{-1}{784} \sum_{i=1}^{784} Y[i] \log (Z[i])  + (1{-} Y[i]) \log(1 {-} Z[i]),
\end{align*}
which, for a fixed $Y$, corresponds to minimizing the average KL-divergence between corresponding pixels that are each treated as a Bernoulli distribution.
Thus, the privacy objective is to conceal the digit, while the utility objective is to minimize (average pixel-level) image distortion.

The mechanism and adversary networks both use two hidden layers with 1000 nodes each and fully-connected links between all layers.
The hidden layers use $\tanh$ as the activation function.
The mechanism input layer uses $784 + 20$ nodes for the image concatenated with 20 random Uniform$[-1,1]$ seed noise values.
The mechanism output layer uses 784 nodes with the sigmoid activation function to directly produce an image in $[0,1]^{784}$.
Note that the mechanism network is an example of the universal approximator architecture mentioned in Section~\ref{sec:universal_approx}.
The attacker input layer uses 784 nodes to receive the image produced by the mechanism.
The attacker output layer uses 10 nodes normalized with a softmax activation function to produce a distribution over the digit labels $\{0, \ldots, 9\}$. We focus on a particular digit and the corresponding release generated by PPAN in \edits{Figure~\ref{fig:mnist_4}}. PPAN learns to add noise at strategic pixels so as to best confound the digit. The third panel shows that adding random noise to each pixel, while keeping the total amount of noise added the same, is not effective at concealing the digit.

\begin{figure}
\centering
    \includegraphics[scale=0.45]{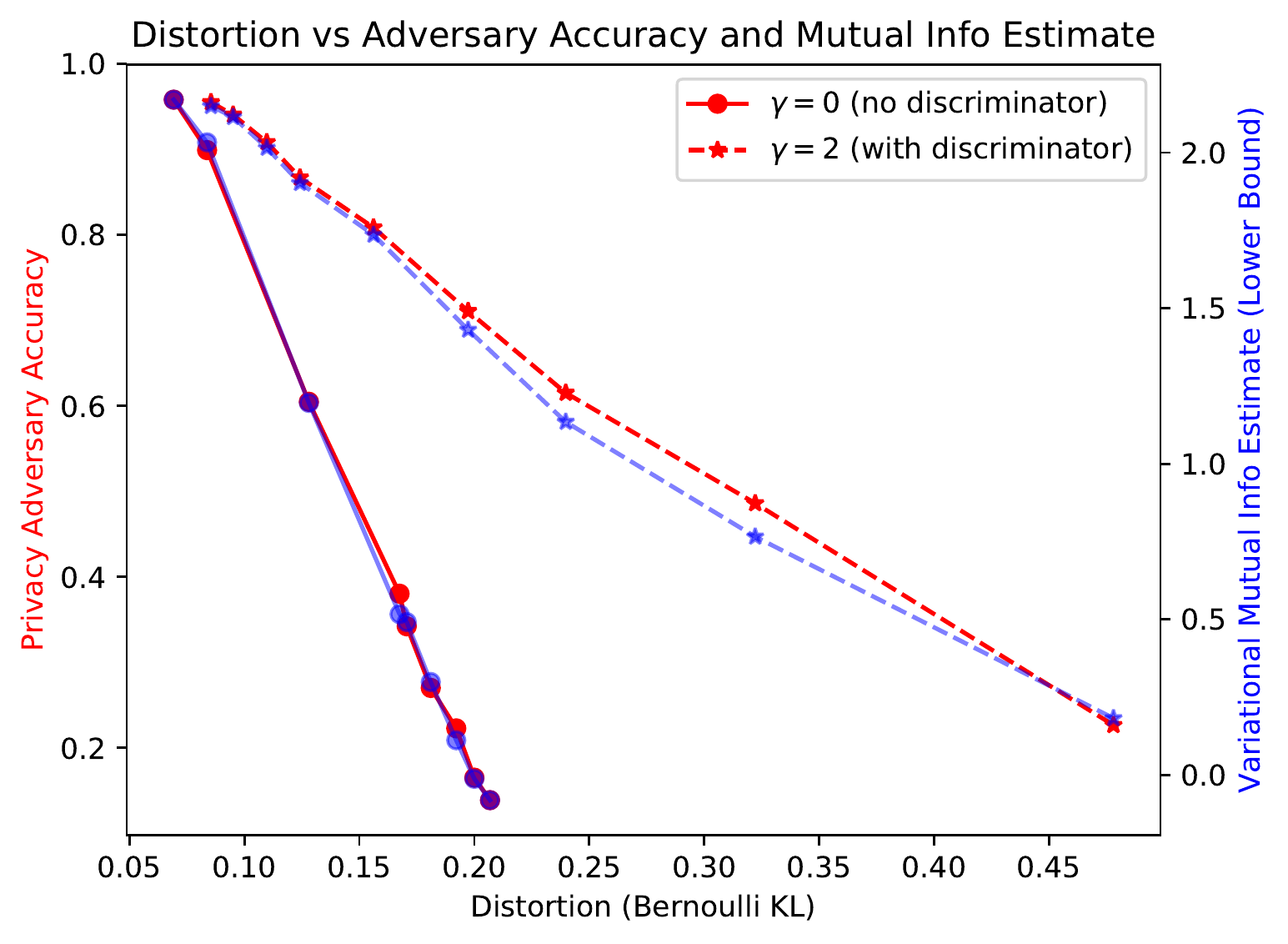}
    \caption{Evaluation of the distortion vs privacy tradeoffs for PPAN applied to the MNIST test set, with privacy measured by adversary accuracy (in red) and estimated mutual information (in blue).}
    \label{fig:MNIST_tradoff}
\end{figure}

\begin{figure}
\centering
\begin{subfigure}[c]{0.35\textwidth}
    \centering
    \includegraphics[scale=0.45]{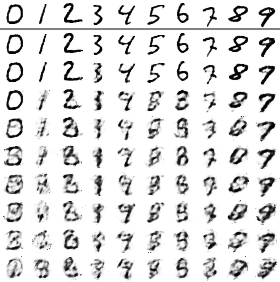}
    \caption{Without additional discriminator (i.e., $\gamma = 0$).
}\label{fig:MNIST_gamma0}
\end{subfigure}
\begin{subfigure}[c]{0.01\textwidth}
    \centering
    \includegraphics[scale=0.12]{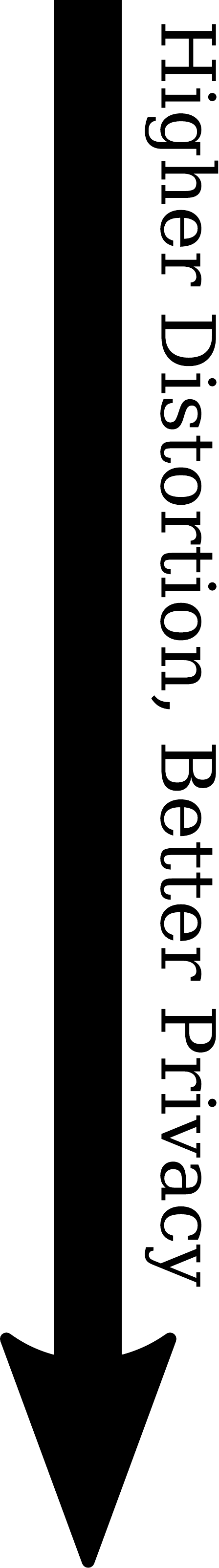}
\end{subfigure}
\begin{subfigure}[c]{0.35\textwidth}
    \centering
    \includegraphics[scale=0.45]{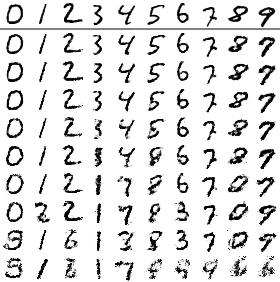}
    \caption{With additional discriminator ($\gamma = 2$).
}\label{fig:MNIST_gamma2}
\end{subfigure}
\begin{subfigure}[c]{0.01\textwidth}
    \centering
    \includegraphics[scale=0.12]{arrow.pdf}
\end{subfigure}
\caption{Examples from applying PPAN to conceal MNIST handwritten digits.
Top row consists of the original test set examples input to the mechanism, while other rows show corresponding mechanism outputs at different tradeoff points.} \label{fig:MNIST}
\end{figure}

In a second set of experiments, we employ the standard GAN approach of adding a discriminator network to further encourage the mechanism to produce output images that resemble realistic digits.
The discriminator network architecture uses a single hidden layer with 500 nodes, and has an output layer with one node that uses the sigmoid activation function.
The discriminator network, denoted by $D_\psi$ with parameters $\psi$, attempts to distinguish the outputs of the mechanism network from the original training images.
Its contribution to the overall loss is controlled by a parameter $\gamma \geq 0$ (with zero indicating its absence).
Incorporating this additional network, the training loss terms are given by
\begin{align}
\mathcal{L}^i_{\text{mnist}}(\theta, \phi, \psi) := &\log Q_\phi(x_i|z_i) + \lambda d(y_i,z_i) \nonumber \\
&{}+ \gamma \log D_\psi(z_i) + \gamma \log (1 - D_\psi(y_i)), \label{eq:MNIST_Loss}
\end{align}
where $z_i$ is generated from the input image $w_i = y_i$ by the mechanism network controlled by the parameters $\theta$.
The overall adversarial optimization objective with both the privacy adversary and the discriminator is given by
\begin{align*}
\min_\theta \max_{\phi, \psi} \frac{1}{n}\sum_{i=1}^n \mathcal{L}^i_{\text{mnist}}(\theta, \phi, \psi).
\end{align*}

We used the 10K test images to objectively evaluate the performance of the trained mechanisms for Figure~\ref{fig:MNIST_tradoff}, which depicts image distortion versus privacy measured by the accuracy of the adversary in recognizing the original digit and the variational lower bound for mutual information obtained by using the posterior distribution of the sensitive attribute learnt by the adversary in~\eqref{eq:MI_var_lb}.

Figure~\ref{fig:MNIST} shows example results from applying trained privacy mechanisms to MNIST test set examples.
The first row depicts the original test set examples input to the mechanism, while the remaining rows each depict the corresponding outputs from a mechanism trained with different values for $\lambda$.
From the second to last rows, the value of $\lambda$ is decreased (from 35 to 8), reducing the emphasis on minimizing distortion.
We see that the outputs start from accurate reconstructions and become progressively more distorted while the digit becomes more difficult to correctly recognize as $\lambda$ decreases.
\edits{Figure~\ref{fig:MNIST_gamma0}} shows the results with the standard PPAN formulation, trained via~\eqref{eq:MNIST_Loss} with $\gamma = 0$, where
we see that the mechanism seems to learn to minimize distortion while rendering the digit unrecognizable, which in some cases results in an output that resembles a different digit.
\edits{Figure~\ref{fig:MNIST_gamma2}} shows the results for the second set of experiments when the additional discriminator network is introduced, which is jointly trained via~\eqref{eq:MNIST_Loss} with $\gamma = 2$.
There we see that the additional discriminator network encourages outputs that more cleanly resemble actual digits, which required lower values for $\lambda$ (ranging from 15 to 2)
to generate distorted images and also led to a more abrupt shift toward rendering a different digit.
For both sets of experiments, the networks were each alternatingly updated once per batch (of 100 images) over 50 epochs of the 60K MNIST training set images.

%% file: analytical_PU.tex
\section{Optimum Privacy Utility Tradeoff for Gaussian Attributes}
\label{sec:PU}
In Section~\ref{sec:results} we compare the (privacy, distortion)
pairs achieved by the model-agnostic PPAN mechanism with the optimal
model-aware privacy-utility tradeoff curve. For jointly Gaussian
attributes and mean squared error distortion, we can obtain, in some
cases, analytical expressions for the optimal tradeoff curve as
described below. Some of the steps in the proofs use bounding
techniques from rate-distortion theory, which is to be expected given
the tractability of the Gaussian model and the choice of mutual
information and mean squared error as the privacy and utility metrics
respectively.

\begin{propn}\label{propn:scalar_OP} (Useful Data only: Scalar
Gaussian with mean squared error)
In problem~\eqref{eq:pp}, let $X,Y$ be jointly Gaussian scalars with
zero means $\mu_X = \mu_Y = 0$, variances $\sigma_X^2, \sigma_Y^2$
respectively, and correlation coefficient $\rho \in [-1,1]$.
Let mean squared error be the distortion measure. If the observation $W=Y$
(Useful Data only observation model), then the optimal release $Z$
corresponding to
\begin{align}\label{eq:op}
\min_{P_{Z|Y}} I(X;Z), \text{ s.t. } \mathbb{E}(Y-Z)^2
\leq \delta \text{ and }  X \leftrightarrow Y \leftrightarrow Z
\end{align}
is given by
\begin{equation*}
Z = 
\begin{cases}
0, & \text{if } \delta \geq \sigma_Y^2 \\
(1 - \delta/\sigma_Y^2)Y + U, & \text{if } \delta <
\sigma_Y^2
\end{cases}
\end{equation*}
where $U \independent (X,Y)$ and $U \sim \mathcal{N}(0,
\delta(1-\delta/\sigma_Y^2))$.
The mutual information leakage caused by releasing $Z$ is
\begin{align*}
I(X;Z) = \max \left\lbrace
0,\frac{1}{2}\log\left(\frac{1}{1-\rho^2+\rho^2\delta/\sigma_Y^2}\right)\right\rbrace.
\end{align*}
\end{propn}

The result of Proposition~\ref{propn:scalar_OP} is known in the
existing literature, e.g.,~\cite{RebolloFD-TKDE10-tCloseToPRAMviaIT} (see Eq.~8)
and~\cite{SankarRP-TIFS13-UtilityPrivacy} (see Example~2). For completeness,
we present the proof of this
result in Appendix~\ref{app:scalar_OP}. The theoretical tradeoff curve
in Figure~\ref{fig:scalar_FD_OP} was obtained using the expressions in
Proposition~\ref{propn:scalar_OP}.

%%%%%%%%%%%%%%%%%%%%%%%%%%%%%%%%%%%%%

The case of Useful Data only observation model for jointly Gaussian {\it
  vector} attributes and mean squared error is also considered in
\cite{RebolloFD-TKDE10-tCloseToPRAMviaIT}, where they provide a
numerical procedure to evaluate the tradeoff curve. Here, we focus on a
special case where we can compute the solution analytically.

Consider the generalization to vector variables of 
\eqref{eq:op}
\begin{align}
\min_{P_{Z\mid Y}} I(X;Z) \text{ such that } &\mathbb{E}(Y-Z)^T(Y-Z)
\leq \delta \nonumber \\
\text{and } &X\leftrightarrow Y\leftrightarrow Z.
\label{eq:op_vector}
\end{align}

Let $X,Y$ be jointly Gaussian vectors of dimensions $m$ and $n$
respectively. We assume that $X,Y$ have zero means $\mu_X = \mu_Y = 0$
and non-singular covariance matrices $\Sigma_X, \Sigma_Y \succ 0$. Let
$\Sigma_{XY}$ denote the cross-covariance matrix and $P :=
\Sigma_X^{-\frac{1}{2}}\Sigma_{XY} \Sigma_Y^{-\frac{1}{2}}$ the
normalized cross-covariance matrix with singular value decomposition
$P = U_P\Lambda_P V_P^\top $. We assume that all singular values of
$P$, denoted by $\rho_i', i = 1, \ldots, \min\{m,n\}$, are strictly
positive.
If
\begin{equation*}
X' := U_P^\top \Sigma_X^{-\frac{1}{2}}X, \;\;\;
Y' := V_P^\top \Sigma_Y^{-\frac{1}{2}}Y, \;\;\; \text{and }
Z' := V_P^\top \Sigma_Y^{-\frac{1}{2}}Z
\end{equation*}
denote reparameterized variables, then $X',Y'$ are zero-mean, jointly
Gaussian, with identity covariance matrices $I_m, I_n$ respectively
and $m \times n$ diagonal cross-covariance matrix $\Lambda_P$. Since
the transformation from $(X,Z)$ to $(X',Z')$ is invertible, $I(X';Z')
= I(X;Z)$. The mean squared error between $Y, Z$:
\begin{equation*}
\mathbb{E}\left[(Y{-}Z)^\top(Y{-}Z)\right] 
= \mathbb{E}\left[(Y'{-}Z')^\top(V_P^\top \Sigma_Y
  V_P)(Y'{-}Z')\right].
\end{equation*}
For the special case when $V_P^\top \Sigma_Y V_P = cI_n$ for some $c >
0$, the vector problem~\eqref{eq:op_vector} reduces to the following
problem:
\begin{align}
\min_{P_{Z'\mid Y'}} I(X';Z') \text{ such that }
&\mathbb{E}(Y'-Z')^T(Y'-Z') \leq \delta/c \nonumber \\
\text{ and }
&X'\leftrightarrow Y'\leftrightarrow Z'.
\label{eq:PU_vec_case1}
\end{align}
\begin{propn}\label{propn:vector_OP}
If 
$\begin{bsmallmatrix} X' \\
Y'\end{bsmallmatrix} \sim \mathcal{N}\Big(\begin{bsmallmatrix}\bm{0}_m \\
\bm{0}_n \end{bsmallmatrix}, \begin{bsmallmatrix}I_m & \Lambda_P \\
\Lambda_P^T & I_n\end{bsmallmatrix}\Big)$, 
then the minimizer of~\eqref{eq:PU_vec_case1} is given by
\begin{equation*}
Z_i' = (1 - \delta_i')Y_i' + U_i, i = 1, \ldots, \min\{m,n\},
\end{equation*}
where $(U_1,\ldots,U_{\min\{m,n\}}) \independent (X',Y')$ and for all
$i$, $U_i \sim \mathcal{N}(0,\delta_i'(1-\delta_i'))$, $\delta_i' :=
\min\{1, t - ({\rho_i'}^{-2} - 1)\}$, where $\rho_i' > 0$ denotes the
$i$-th main diagonal entry of $\Lambda_P$, and the value of parameter
$t$ can be found by the equation
$\sum_i \delta_i' = \delta/c$. The mutual information between the
release and the sensitive attribute is
$I(X', Z') = \sum_{i=1}^{\min\{m,n\}} \max\{0, -0.5 \log(1 -
{\rho_i'}^2 + {\rho_i'}^2 \delta_i')\}$.
\end{propn}
The proof of the above proposition is given in Appendix~\ref{app:vector_OP}.
We evaluate the above parametric expression for
various values of $\delta$ in order to obtain the theoretical tradeoff
curves in Figure~\ref{fig:vector_OP}.

For the case of full data observation, we have the following result.
%%%%%%%%%%%%%%%%%%%%%%%%%%%%%%%%%%%%%
\begin{propn}\label{propn:scalar_FD}
%%%%%%%%%%%%%%%%%%%%%%%%%%%%%%%%%%%%%
(Full Data: Scalar Gaussian with mean squared error)
In problem~\eqref{eq:pp}, let $X,Y$ be jointly Gaussian scalars with
zero means, unit variances,
and correlation coefficient $\rho \in [0,1]$.
Let mean squared error be the distortion measure. If the observation
$W=(X,Y)$ (full data observation model), then the optimal release $Z$
corresponding to
\begin{align}
\min_{P_{Z|X,Y}} I(X;Z), \quad \text{such that} \quad
\mathbb{E}(Y-Z)^2 \leq \delta
\label{eq:fd}
\end{align}
is given by
\begin{equation*}
Z = (1-\delta)Y - (X-\rho Y)\sqrt{\frac{\delta(1-\delta)}{1-\rho^2}}.
\end{equation*}
The mutual information leakage caused by this release is 
$I(X;Z) = 0$ if $\delta \geq \rho^2$, and if $\delta < \rho^2$:
\begin{equation*}
I(X;Z) = \frac{1}{2}\log \left( \frac{1}{1-
  \left(\sqrt{\rho^2(1-\delta)} -
  \sqrt{(1-\rho^2)\delta}\right)^2}\right).
\end{equation*}
\end{propn}
The proof of the above proposition is presented in
Appendix~\ref{app:scalar_FD}. The theoretical tradeoff curve in
Figure~\ref{fig:scalar_FD_OP} was obtained using the above expression.

%% file: conclusion.tex
\section{Conclusion}

In this work we introduced and developed a practical, data-driven method for optimizing privacy-preserving data release mechanisms within the well-established information-theoretic framework.
The key to this approach is the application of adversarially-trained neural networks, where the mechanism is realized as a randomized network, and a second network acts as a privacy adversary that attempts to recover sensitive information.
By estimating the posterior distribution of the sensitive variable given the released data, the adversarial network enables a variational approximation of mutual information.
This allows our method to approach the information-theoretically optimal privacy-utility tradeoffs, which we demonstrate in experiments with discrete and continuous synthetic data.
We also conducted experiments with the MNIST handwritten digits dataset, where we trained a mechanism that trades off between minimizing the pixel-level image distortion and concealing the digit.

%% file: algorithm.tex
\section{Algorithm Pseudocode} \label{sec:pseudocode}

The high-level training procedure is the typical iterative network updates by gradient descent with loss gradients estimated over rotating mini-batches of training data for a given number of epochs.
For each batch of training data $\{(w_i, x_i, y_i)\}_{i=1}^n$, the networks are updated as follows:
\begin{enumerate}
\item For each $i \in \{1, \ldots, n\}$, sample $k$ independent releases $\{z_{i,j}\}_{j=1}^k \stackrel{\text{iid}}{\sim} P_\theta(z|w_i)$.
See Sections~\ref{sec:sampling} and~\ref{sec:GMM} for specific sampling techniques.
Note that in our experiments, we used $k = 1$, and denote the $n$ total samples by $\{z_i\}_{i=1}^n$.
This step is skipped for the finite alphabet case of Section~\ref{sec:finite_arch}.
\item Compute the objective
\[
\mathcal{L} := \frac{1}{n}\sum_{i=1}^n \mathcal{L}^i(\theta, \phi),
\]
where $\mathcal{L}^i$ is given by either~\eqref{eq:PPAN_loss},~\eqref{eqn:finite_arch},~\eqref{eqn:gauss-loss},~\eqref{eq:MNIST_Loss}, or~\eqref{eqn:gmm-loss} depending on the specific experiment and sampling method.
\item Update the adversarial network parameters $\phi$ by ascending the gradient $\nabla_\phi \mathcal{L}$.
\item For the MNIST experiment (see Section~\ref{sec:mnist} and~\eqref{eq:MNIST_Loss}), if an additional discriminator network is used (i.e., $\gamma > 0$), update the discriminator network parameters $\psi$ by ascending the gradient $\nabla_\psi \mathcal{L}$.
\item Recompute the objective $\mathcal{L}$ and update the release mechanism network parameters $\theta$ by descending the gradient $\nabla_\theta \mathcal{L}$.
\end{enumerate}

%% file: appendix.tex
\section{Proofs of Propositions} \label{sec:proofs}
\subsection{Proof of Proposition~\ref{propn:scalar_OP}}\label{app:scalar_OP}
\begin{proof}
We can expand the mutual information term as follows,
\begin{align}
I(X;Z) &= h(X) - h(X| Z), \nonumber \\
       &= h(X) - h(X-\rho\sigma_X Z/\sigma_Y| Z), \nonumber \\
       &\geq h(X) - h(X - \rho\sigma_X Z/\sigma_Y),\label{eq:MI_cond}\\
       &\geq 0.5 \log 2\pi e\sigma_X^2 - h\left(\mathcal{N}(0,
\mathbb{E}[(X - \rho\sigma_X Z/\sigma_Y)^2]\right),\label{eq:MI_normal}\\
       &=
\frac{1}{2}\log\left(\frac{\sigma_X^2}{\mathbb{E}[(X-\rho\sigma_X
    Z/\sigma_Y)^2]}\right)
\label{eq:MI}.
\end{align}
Inequality~\eqref{eq:MI_cond} is true because conditioning can only
reduce entropy and inequality~\eqref{eq:MI_normal} is true since
the zero-mean normal distribution has the maximum
entropy for a given value of the second moment.  Let $T := X -
\rho\sigma_XY/\sigma_Y$, then $T$ is jointly Gaussian and we have that
\begin{align*}
&\text{Cov}\left(
\begin{bmatrix}
T \\ 
Y 
\end{bmatrix}
\right)\\
&=
\begin{bmatrix}
1 & -\rho\sigma_X/\sigma_Y\\
0 & 1
\end{bmatrix}
\begin{bmatrix}
\sigma_X^2 & \rho \sigma_X\sigma_Y\\
\rho\sigma_X\sigma_Y & \sigma_Y^2
\end{bmatrix}
\begin{bmatrix}
1 & 0\\
-\rho\sigma_X/\sigma_Y & 1
\end{bmatrix}\\
&=
\begin{bmatrix}
\sigma_X^2(1 - \rho^2) & 0\\
0 & \sigma_Y^2
\end{bmatrix}.
\end{align*}
Hence, $T$ is independent of $Y$. Since $X \leftrightarrow Y \leftrightarrow Z$
also forms a Markov chain, we have that $T$ is
conditionally independent of $Z$ given $Y$.  Due to the distortion
constraint, we can upper bound $\mathbb{E}[(X-\rho\sigma_X
  Z/\sigma_Y)^2]$ in the following manner.
\begin{align}
&\mathbb{E}\Big[\left(X-\frac{\rho\sigma_X}{\sigma_Y} Z\right)^2\Big] \nonumber \\
&= \sigma_X^2 + \frac{\rho^2\sigma_X^2}{\sigma_Y^2}\mathbb{E}[Z^2] - 2
\frac{\rho\sigma_X}{\sigma_Y} \mathbb{E}\left[\left(T +
  \frac{\rho\sigma_X}{\sigma_Y}Y\right)Z\right],\nonumber \\
&\leq \sigma_X^2 + \frac{\rho^2\sigma_X^2}{\sigma_Y^2}(\delta -
\sigma_Y^2 + 2\mathbb{E}[YZ]) \nonumber \\
&\qquad \qquad- 2\frac{\rho\sigma_X}{\sigma_Y}
\left(\frac{\rho\sigma_X}{\sigma_Y}\mathbb{E}[YZ] +
  \mathbb{E}[TZ]\right),
\label{eq:distortion} \\
&= \sigma_X^2(1 - \rho^2) +
\rho^2\delta\sigma_X^2/\sigma_Y^2.\label{eq:WZgY}
\end{align}
Inequality~\eqref{eq:distortion} is true because $\mathbb{E}[(Y-Z)^2]\leq
\delta$, and equation~\eqref{eq:WZgY} is true because
\begin{align*}
\mathbb{E}[TZ] = \mathbb{E}_Y[\mathbb{E}_{T,Z| Y}[TZ]]
&\overset{\text{(i)}}{=} \mathbb{E}_Y\left[\mathbb{E}_{T|
    Y}[T]\mathbb{E}_{Z| Y}[Z]\right]\\
&\overset{\text{(ii)}}{=}
\mathbb{E}_Y\left[\mathbb{E}_{T}[T]\mathbb{E}_{Z| Y}[Z]\right]
\overset{\text{(iii)}}{=} 0,
\end{align*}
where (i) is true because $T\independent Z \mid Y$, (ii) is true
because $T \independent Y$ and (iii) is true because $T$ has zero
mean. Thus by equations~\eqref{eq:MI} and~\eqref{eq:WZgY}, we get that
\begin{align}
\min_{\substack{X\leftrightarrow Y \leftrightarrow Z,\\ \mathbb{E}[(Y-Z)^2] \leq
  \delta}} I(X;Z) \geq \max \left\lbrace
0,\frac{1}{2}\log\left(\frac{1}{1-\rho^2+\rho^2\delta/\sigma_Y^2}\right)\right\rbrace.\label{eq:MI_lb}
\end{align}
For the choice of $Z$ as stated in the proposition, we can check that
$X$ and $Z$ are jointly Gaussian with $I(X;Z) =
0.5\log_2(\sigma_X^2/\text{Var}(X | Z))$ and $\text{Var}(X | Z)
= \sigma_X^2(1 - \rho^2 + \rho^2\delta/\sigma_Y^2)$. Thus $Z$ attains
the lower bound for the privacy-leakage in~\eqref{eq:MI_lb}
when $\delta < \sigma_Y^2$. Otherwise, the lower bound on mutual
information is $0$ and can be attained by $Z=0$.
\end{proof}

\subsection{Proof of Proposition~\ref{propn:vector_OP}}\label{app:vector_OP}
\begin{proof}
(a) If $m \leq n$, then $(X_1',Y_1'), \ldots, (X_m',Y_m'),
  Y_{m+1}',\allowbreak \ldots, Y_n'$ are independent because they are jointly
  Gaussian and for all $i\neq j$, ${\rm Cov}(X_i',X_j') = {\rm
    Cov}(X_i',Y_j') = {\rm Cov}(Y_i',Y_j') = 0$. Similarly, if $m \geq
  n$, then $(X_1',Y_1'), \ldots,\allowbreak (X_n',Y_n'),\allowbreak 
  X_{n+1}',\allowbreak \ldots,\allowbreak X_m'$
  are independent.

In the following, we use the following well-known properties of
mutual information and conditional mutual information. For any three
random variables $A,B,C$, (b) $I(A;B) = I(B;A) \geq 0$, (c) $I(A;B)=0
\Leftrightarrow A \independent B$, (d) $I(A;C|B) \geq 0$, (e)
$I(A;C|B) = 0 \Leftrightarrow (A\independent C)|B$, (f) $I(A;B,C) =
I(A;B) + I(A;C|B)$ so that $I(A;B,C) \geq I(A;B)$.

If $m \leq n$, then 
%$I(X';Z') =$ $I(X_1',\ldots, X_m'; Z')$
%$\overset{\text{(f)}}{=}$ $\sum_{i=1}^m I(X_i;Z'|X_1, \ldots, X_{i-1})$
%$\overset{\text{(f,a,c)}}{=}$ $\sum_{i=1}^m I(X_i;Z', X_1, \ldots, X_{i-1})$
%$\overset{\text{(f)}}{\geq}$ $\sum_{i=1}^m I(X_i;Z')$ $\overset{\text{(f)}}{\geq}$
%$\sum_{i=1}^m I(X_i;Z_i)$. 
\begin{align*}
I(X';Z') &= I(X_1',\ldots, X_m'; Z')\overset{\text{(f)}}{=}\sum_{i=1}^m I(X_i;Z'|X_1, \ldots, X_{i-1})\\
&\overset{\text{(f,a,c)}}{=}\sum_{i=1}^m I(X_i;Z', X_1, \ldots, X_{i-1})\\
&\overset{\text{(f)}}{\geq}\sum_{i=1}^m I(X_i;Z')\overset{\text{(f)}}{\geq}\sum_{i=1}^m I(X_i;Z_i).
\end{align*}
Similarly, if $m \geq n$, $I(X';Z')$ $\geq$
$\sum_{i=1}^m I(X_i;Z')$ $\geq$ $\sum_{i=1}^n I(X_i;Z_i)$. 
Thus in general,
\begin{equation*}
I(X';Z') \geq \sum_{i=1}^{\min\{m,n\}} I(X_i;Z_i).
\end{equation*}

The distortion constraint in \eqref{eq:PU_vec_case1} implies that
$\sum_{i=1}^{\min\{m,n\}}\mathbb{E}[(Y_i'-Z_i')^2] \leq \delta/c$. Thus
the optimal function value in \eqref{eq:PU_vec_case1} is lower bounded
by the optimum value of the following problem.
%
%\begin{align}
%%
%&\min_{P_{Z'\mid Y'}} \sum_{i=1}^{\min\{m,n\}}I(X_i';Z_i') \quad
%  \text{s.t.}\quad
%  \sum_{i=1}^{\min\{m,n\}}\mathbb{E}[(Y_i'-Z_i')^2]\leq \delta/c
%  \quad\text{and}\quad X'\leftrightarrow Y'\leftrightarrow
%  Z'. \nonumber\\
%\equiv &\min_{\sum \delta_i'\leq \delta/c, P_{Z'\mid Y'}}
%\sum_{i=1}^{\min\{m,n\}} I(X_i';Z_i') \quad\text{s.t.}\quad \forall i,
%\mathbb{E}[(Y_i'-Z_i')^2] \leq \delta_i' \quad\text{and}\quad
%X'\leftrightarrow Y' \leftrightarrow Z'. \label{eq:PU_vec_case1b}
%%
%\end{align}
\begin{IEEEeqnarray}{RtL}
\min_{P_{Z'\mid Y'}} \sum_{i=1}^{\min\{m,n\}}I(X_i';Z_i')
& s.t. & \!\!\!\sum_{i=1}^{\min\{m,n\}}\mathbb{E}[(Y_i'{-}Z_i')^2]\leq \frac{\delta}{c} \IEEEnonumber\\
& and & X'\leftrightarrow Y'\leftrightarrow Z'. \IEEEnonumber\\
\equiv \min_{\substack{\sum \delta_i'\leq \delta/c,\\ P_{Z'\mid Y'}}}
\sum_{i=1}^{\min\{m,n\}} I(X_i';Z_i')
& s.t. & \forall i, \mathbb{E}[(Y_i'-Z_i')^2] \leq \delta_i' \IEEEnonumber\\
& and & X'\leftrightarrow Y' \leftrightarrow Z'. \label{eq:PU_vec_case1b}
\end{IEEEeqnarray}

Let $Y_{\sim i} := \{Y_1,\ldots, Y_n\} \backslash Y_{i}$. Since
$X'\leftrightarrow Y' \leftrightarrow Z'$ forms a Markov chain, if $m\leq n$, we have 
%$0
%\overset{\text{(e)}}{=}$ $I(X';Z'|Y') = $ $I(X_1,\ldots,X_m;Z'|Y_1,\ldots,
%Y_n)$ $\overset{\text{(f)}}{=}$ $\sum_{i=1}^{m} I(X_i;Z'|Y_1,\ldots, Y_n,
%X_1, \ldots, X_{i-1})$ $\overset{\text{(f,a,c)}}{=}$ $\sum_{i=1}^{m}
%I(X_i;Z', X_1, \ldots, X_{i-1}, Y_{\sim i} | Y_i)$ $\overset{\text{(f)}}{\geq}$ $\sum_{i=1}^{m} I(X_i;Z_i|Y_i)$
%$\overset{\text{(d)}}{\geq} 0$. 
\begin{align*}
0 &\overset{\text{(e)}}{=} I(X';Z'|Y') = I(X_1,\ldots,X_m;Z'|Y_1,\ldots,
Y_n) \\
&\overset{\text{(f)}}{=} \sum_{i=1}^{m} I(X_i;Z'|Y_1,\ldots, Y_n,
X_1, \ldots, X_{i-1})\\
&\overset{\text{(f,a,c)}}{=}\sum_{i=1}^{m}
I(X_i;Z', X_1, \ldots, X_{i-1}, Y_{\sim i} | Y_i)\\
&\overset{\text{(f)}}{\geq}\sum_{i=1}^{m} I(X_i;Z_i|Y_i)\overset{\text{(d)}}{\geq} 0.
\end{align*}
Thus,
\begin{equation*}
0 \geq \sum_{i=1}^{m} I(X_i;Z_i|Y_i) \geq 0.
\end{equation*}
A similar expression can be derived for the case $m \geq n$. In
general, for all $i = 1, \ldots, \min\{m,n\}$, $X_i \leftrightarrow Y_i \leftrightarrow Z_i$ forms a
Markov chain. Thus for output perturbation, the Markov constraint on
the vectors passes through as a Markov constraint on the individual
components of the variables. We can therefore rewrite problem
\eqref{eq:PU_vec_case1b} as follows,
\begin{align*}
\min_{\sum \delta_i' \leq \delta/c} \sum_{i=1}^{\min\{m,n\}}
\min_{P_{Z_i'\mid Y_i'}}I(X_i';Z_i'), &\text{ s.t. } \forall i,
\mathbb{E}[(Y_i'-Z_i')^2] \leq \delta_i'\\
&\text{and }X_i'\leftrightarrow Y_i' \leftrightarrow Z_i'.
\end{align*}
For each $i$, the solution to the inner constrained minimization
problem is given by Proposition~\ref{propn:scalar_OP}. Plugging in the
solution we arrive at the following constrained convex minimization
problem
\begin{align*}
\min_{\sum \delta_i' \leq \delta/c} \sum_{i=1}^{\min\{m,n\}}
\max\left\{0,\frac{1}{2}\log \left(\frac{1}{1-{\rho_i'}^2 +
  {\rho_i'}^2\delta_i'}\right)\right\}
\end{align*}
where $\rho_i' = \mathbb{E}[X_i'Y_i']$ and we have used the expression
for the optimal privacy-leakage in the scalar case, i.e.,
Eq.~\eqref{eq:MI_lb} with $\sigma_Y^2=1$.
The Lagrangian of the above convex program has the following form
\begin{align*}
  \mathcal{L}(\bm{\delta}', \bm{\eta}, \zeta) :=&
  \sum_{i=1}^{\min\{m,n\}}\frac{1}{2}\log \left(\frac{1}{1-{\rho_i'}^2
    + {\rho_i'}^2\delta_i'}\right)\\
    &{} + \zeta\left(\sum_{i=1}^{\min\{m,n\}}\delta_i' -
  \frac{\delta}{c}\right) + \sum_{i=1}^{\min\{m,n\}}\eta_i(\delta_i' -
  1),
\end{align*} 
where $\bm{\delta}' = (\delta_1',\ldots,\delta_{\min\{m,n\}}')$,
$\bm{\eta} = (\eta_1,\ldots,\eta_{\min\{m,n\}})$, and $\zeta$ and all
the $\eta_i$'s are non-negative Lagrange multipliers. Here, the
non-negativity condition associated with $\max\{0,\cdot\}$ has been
subsumed by requirement that $\delta_i' \leq 1$ for all $i$. The
Karush-Kuhn-Tucker (KKT) conditions for optimality are as follows,
\begin{align*}
%
%\sum_{i=1}^{\min\{m,n\}} \delta_i' = \frac{\delta}{c},
%
\text{For each }i,\quad &0 \leq \delta_i' \leq 1, 
\eta_i \geq 0, 
\eta_i(\delta_i' - 1) = 0,\\
&\frac{\partial \mathcal{L}}{\partial \delta_i'}=0 \Rightarrow \eta_i =
\frac{1}{2(\delta_i'-1+{\rho_i'}^{-2})} - \zeta,
\end{align*}
and $\sum_{i=1}^{\min\{m,n\}} \delta_i' = \delta/c$. 
This implies that if for any $i$,
\begin{align*}
(2\zeta)^{-1} > {\rho_i'}^{-2} \Leftrightarrow \eta_i > 0, \text{ then
} &\delta_i' = 1,\\
\text{ otherwise } &\delta_i' = (2\zeta)^{-1} -({\rho_i'}^{-2} - 1).
\end{align*}
The value of $(2\zeta)^{-1}$ can be found by the equation 
\begin{equation*}
\sum_{i=1}^{\min\{m,n\}}\max\left\lbrace 0, \min\{1, (2\zeta)^{-1} -
({\rho_i'}^{-2} - 1)\}\right\rbrace = \frac{\delta}{c},
\end{equation*}
which is a modified water-filling solution. Based on the value of
$\delta_i'$, we can construct a $Z'_i$ that attains the lower bound on
the mutual information by setting $\sigma_Y^2 = 1$ in the results of
Proposition~\eqref{propn:scalar_OP}.
\end{proof}

%%%%%%%%%%%%%%%%%%%%%%%%%%%%%%%%%%%%%%%%%%%%%%%%%%%%%%%%%%%%%%%%%%%%%%%%%%%%
\subsection{Proof of Proposition \ref{propn:scalar_FD}}\label{app:scalar_FD}
\begin{proof}
In this proposition, $X$ and $Y$ are assumed to be jointly Gaussian
with zero means, unit variances, and correlation coefficient $\rho \in
[0,1]$. Consider the Linear Minimum Mean Squared Error (LMMSE)
estimate of $X$ given $Z$ denoted as $\widehat{\mathbb{E}}[X| Z] =
\mathbb{E}[XZ]Z/\mathbb{E}[Z^2]$. Then, similar to the proof of
Proposition~\ref{propn:scalar_OP}, we can expand the mutual
information in the following manner.
\begin{align*}
I(X;Z) &= h(X) - h(X | Z) = h(X) - h(X - \widehat{\mathbb{E}}[X|
Z] | Z)\\
&\geq h(X) - h(X - \widehat{\mathbb{E}}[X| Z])\\
&\geq h(X) -
h\Big(\mathcal{N}\Big(0, \mathbb{E}\Big[\left(X -
  \widehat{\mathbb{E}}[X| Z]\right)^2\Big]\Big)\Big)\\
&= -\frac{1}{2}\log \left(1 -
  \frac{(\mathbb{E}[XZ])^2}{\mathbb{E}[Z^2]}\right),
\end{align*}
where in writing the last equality we have used the fact that
$\mathbb{E}\left[\widehat{\mathbb{E}}[X| Z](X -
  \widehat{\mathbb{E}}[X| Z])\right] = 0$ by the orthogonality
principle of least squares estimation. Thus, we have that
\begin{equation}
\min_{\mathbb{E}[(Y-Z)^2] \leq \delta} I(X;Z)\geq
-\frac{1}{2}\log\left[1 - \min_{\mathbb{E}[(Y-Z)^2] \leq
    \delta}\frac{(\mathbb{E}[XZ])^2}{\mathbb{E}[Z^2]}\right]
\label{eq:PU}
\end{equation}
Below, we focus on the minimization problem on the right side of
Eq.~\eqref{eq:PU}. It will turn out that for the minimizing $Z^\star$,
we will have equality in Eq.~\eqref{eq:PU}.
In what follows, it is helpful to think of the random variables $X, Y,
Z$ as vectors in the vector space $\mathcal{L}_2$ of all random
variables with finite second moments over the underlying probability
space. We will emphasize the vector nature by denoting the random
variables $X,Y,Z$ by their corresponding bold lowercase letters
$\bm{x}, \bm{y}, \bm{z}$ respectively. The expectation operator on the
product of two random variables in $\mathcal{L}_2$ is an inner
product, and hence we can write the optimization problem of interest
as follows,
\begin{align}
\bm{z}^\star := 
\arg\min_{\bm{z}: \lVert \bm{z}-\bm{y} \rVert^2 \leq \delta}
\frac{|\langle\bm{x}, \bm{z} \rangle|^2}{\lVert \bm{z} \rVert^2} = 
\arg\min_{\bm{z}: \lVert \bm{z}-\bm{y} \rVert^2 \leq \delta}
|\langle\bm{x}, \frac{\bm{z}}{\lVert \bm{z} \rVert} \rangle|^2,
\label{eq:vecform1}
\end{align}
where, $\lVert\bm{x}\rVert = \lVert\bm{y}\rVert = 1$, and
$\langle\bm{x},\bm{y} \rangle = \rho$.
Let $\hat{\imath} := \bm{x}$,
$\hat{\jmath} := \frac{1}{\sqrt{1-\rho^2}}({\bm y} - \rho {\bm x}) =
\frac{{\bm y} - \text{Proj}_{\text{Span}({\bm x})}({\bm y})}
{||{\bm y} - \text{Proj}_{\text{Span}({\bm x})}({\bm y})||}$,
and $\hat{k} := \frac{{\bm z} -
  \text{Proj}_{\text{Span}(\bm{x},\bm{y})}({\bm z})}{||{\bm z} -
  \text{Proj}_{\text{Span}(\bm{x},\bm{y})}({\bm z})||}$.
Then $\hat{\imath}, \hat{\jmath}, \hat{k}$ are unit vectors along
three orthogonal coordinate axes and $\bm{y} = \rho \hat{\imath} +
\sqrt{1 - \rho^2}\hat{\jmath}$. Let $\bm{t} :=
\bm{z}-\bm{y}=t_1\hat{\imath} + t_2 \hat{\jmath} + t_3\hat{k}$ so that
$\bm{z} = (t_1+\rho)\hat{\imath} + (t_2 + \sqrt{1-\rho^2})\hat{\jmath}
+ t_3\hat{k}$.
Then the problem in Eq.~\eqref{eq:vecform1} is equivalent to the
following one
\begin{align}
(t_1^\star,t_2^\star,t_3^\star) := \underset{\substack{t_1,t_2,t_3:\\ t_1^2 +
    t_2^2+t_3^2 \leq \delta}}{\arg\min} \left[
    \frac{(t_1+\rho)^2}{t_1^2{+}t_2^2{+}t_3^2{+}2t_1\rho{+}2t_2\sqrt{1{-}\rho^2}{+}1}\right].
\label{eq:PU_RVa}
\end{align}
\noindent{{\bf Case} $\rho^2 \leq \delta$:} If $\rho^2 \leq \delta$,
then $t_1^\star = -\rho, t_2^\star = t_3^\star = 0$ is a minimizer of the
problem in~\eqref{eq:PU_RVa} and $\bm{z}^\star = \sqrt{1 -
  \rho^2}\hat{\jmath} = \bm{y} - \rho \bm{x}$. This solution is
displayed along with $\bm{x}$ and $\bm{y}$ in Figure~\ref{fig:FD_case1}
and has an immediate geometric interpretation. One can see that
$\langle \bm{x}, \bm{z} \rangle = 0$ which implies that $X
\independent Z$ because then $Z$ and $X$ are uncorrelated and $Z$,
being a linear combination of $X$ and $Y$, is jointly Gaussian with
them. Also then, $\lVert \bm{z} - \bm{y}\rVert^2 \leq \delta$, or
equivalently, $\mathbb{E}[(Y - Z)^2] \leq \delta$. Thus, the lower
bound of $0$ for $I(X;Z)$ is attained in~\eqref{eq:fd} by this
solution.
\begin{figure}[!htb]
  \centering
  \includegraphics[width=0.3\textwidth]{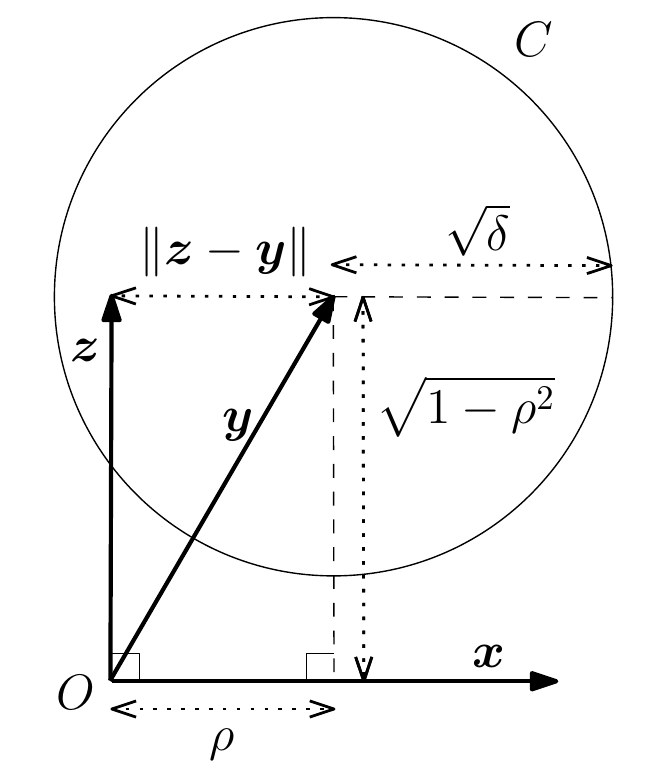}
  \caption{Solution to problem \eqref{eq:fd} for the case when $0 \leq
    \rho \leq +\sqrt{\delta}$. In the figure, $\bm{x} ,\bm{y}$
    represent unit length vectors with inner product equal to $\rho
    \in [0,1]$. The vector $\bm{z}$ is perpendicular to $\bm{x}$ and
    lies within a distortion sphere of radius $\sqrt{\delta}$ around
    $\bm{y}$. The circle $C$ is the projection of the distortion
    sphere on the $\bm{x}$-$\bm{y}$ plane and point $O$ is the
    origin. The dotted lines with hollow arrowheads denote the lengths
    of various quantities.}\label{fig:FD_case1}
\end{figure}

\noindent{{\bf Case} $\rho^2 > \delta$:}
If $(t_1,t_2,t_3)$ is feasible in~\eqref{eq:PU_RVa}, i.e., $t_1^2 +
t_2^2 + t_3^3 \leq \delta$ then so is $(t_1',t_2',t_3') :=
(t_1,+\sqrt{t_2^2+t_3^2},0)$. If $t_3 \neq 0$ then $(t_1',t_2',t_3')$
strictly dominates $(t_1,t_2,t_3)$ because the denominator of the
objective function in~\eqref{eq:PU_RVa} is strictly larger for
$(t_1',t_2',t_3')$ than for $(t_1,t_2,t_3)$. Thus, we must have
%
%\begin{align}
%t_3^\star = 0,
%\label{eq:onxyplane} 
%\end{align}
$t_3^\star = 0$, 
otherwise we can strictly improve (i.e., strictly decrease) the
objective function value contradicting the optimality of
$(t_1^\star,t_2^\star,t_3^\star)$. Geometrically, this means that
$\bm{z}^\star$ must lie in the two dimensional subspace spanned by
$\bm{x}$ and $\bm{y}$. Consequently, the minimization problem in~\eqref{eq:PU_RVa} reduces to
\begin{align}
(t_1^\star,t_2^\star) = \underset{t_1,t_2: t_1^2+t_2^2\leq
    \delta}{\arg\min}\left[\frac{(t_1+\rho)^2}{t_1^2+t_2^2+2t_1\rho+2t_2\sqrt{1-\rho^2}+1}\right].
\label{eq:PU_RVb}
\end{align}
If $(t_1,t_2)$ is feasible in~\eqref{eq:PU_RVb}, i.e., $t_1^2 + t_2^2
\leq \delta$ then so is $(t_1',t_2') :=
(t_1,+\sqrt{t_2^2+(\delta-t_1^2-t_2^2)})$. If $t_1^2 + t_2^2 < \delta$,
then $(t_1',t_2')$ strictly dominates $(t_1,t_2)$ because the
denominator of the objective function in~\eqref{eq:PU_RVb} is
strictly larger for $(t_1',t_2')$ than for $(t_1,t_2)$. Thus, we must
have 
\begin{align}
(t_1^\star)^2 + (t_2^\star)^2 &= \delta,
\label{eq:oncircle}
\end{align}
otherwise we can strictly improve (i.e., strictly decrease) the
objective function value contradicting the optimality of
$(t_1^\star,t_2^\star)$. Geometrically, this means that $\bm{z}^\star$
must lie on the circle of radius $+\sqrt{\delta}$ centered at
$\bm{y}$.

Finally, we observe that if $\bm{z}$ is feasible in~\eqref{eq:vecform1}, i.e., $\lVert \bm{y}-\bm{z} \rVert^2 \leq
\delta$, then so is $\bm{z}' :=
\text{Proj}_{\text{Span}(\bm{z})}(\bm{y}) = \gamma \bm{z}$, where
$\gamma = \frac{\langle\bm{y}, \bm{z} \rangle}{||\bm{z}||^2}$. This is
because the orthogonal projection of a vector onto a subspace is the
vector in the subspace closest to it so that $||\bm{y} - \bm{z}'||^2
\leq ||\bm{y} - \bm{z}||^2$. Also observe that the value of the
objective function in~\eqref{eq:vecform1} is the same for both
$\bm{z}$ and $\gamma \bm{z}$ and that $(\bm{y} - \bm{z}') \perp
\bm{z}'$. Thus, we may assume that there is an optimal solution
$\bm{z}^\star$ such that $(\bm{y}-\bm{z}^\star) \perp \bm{z}^\star$
for if not, we can rescale $\bm{z}^\star$ suitably to ensure this
property without affecting the objective function or violating the
distortion constraint. Since $(\bm{z}^\star - \bm{y}) =
t_1^\star\hat{\imath} + t_2^\star \hat{\jmath}$ and $\bm{y} = \rho
\hat{\imath} + \sqrt{1-\rho^2} \hat{\jmath}$, the orthogonality
condition $(\bm{y}-\bm{z}^\star) \perp \bm{z}^\star$ can be restated
as
\[
t_1^\star (t_1^\star + \rho) + t_2^\star (t_2^\star + \sqrt{1-\rho^2})
= 0
\]
which simplifies to
\begin{align}
(t_1^\star)^2 + (t_2^\star)^2 + t_1^\star \rho + t_2^\star
\sqrt{1-\rho^2} = 0
\label{eq:tangentcondition}
\end{align}
Combining~\eqref{eq:tangentcondition} with~\eqref{eq:oncircle}
we get
\[
\delta + t_1^\star \rho + \sqrt{(\delta - (t_1^\star)^2)(1-\rho^2)} =
0.
\]
This reduces to the following quadratic equation for $t_1^\star$ with
two real roots
\begin{align*}
&(t_1^\star)^2 + 2\delta \rho t_1^\star + \delta^2-\delta(1-\rho^2) = 0\\
&\implies  t_1^\star = -\delta \rho \pm
\sqrt{\delta(1-\delta)(1-\rho^2)}.
\end{align*}
We note that $\delta < 1$ since we are considering the case $\delta <
\rho^2 \leq 1$.
Of the two real roots, $t_1^\star = -\delta \rho -
\sqrt{\delta(1-\delta)(1-\rho^2)}$ has a lower objective value in~\eqref{eq:PU_RVb}.
Using this value for $t_1^\star$ and setting $t_2^\star = \sqrt{\delta
  - (t_1^\star)^2}, t_3^\star = 0$, we can conclude that for the case
when $\delta < \rho^2$, the random variable
\begin{align}
Z^\star := (1-\delta)Y - (X-\rho
Y)\sqrt{\frac{\delta(1-\delta)}{1-\rho^2}}
\label{eq:bestZ}
\end{align} 
attains the lower bound on the mutual information, i.e.,%which equals
\begin{align}
I(X;Z) =
\frac{1}{2}\log \left( \frac{1}{1- \left(\sqrt{\rho^2(1-\delta)} -
  \sqrt{(1-\rho^2)\delta}\right)^2}\right).
\label{eq:FD}
\end{align}
\begin{figure}
  \centering
  \includegraphics[width=0.4\textwidth]{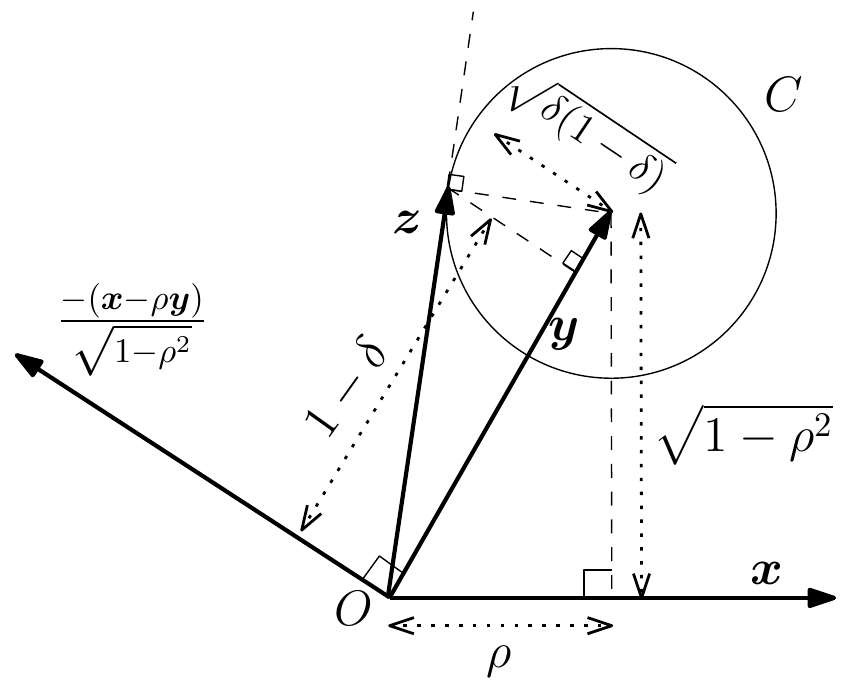}
  \caption{Geometric interpretation of the solution to the problem in~\eqref{eq:PU_RVb} for the case when $\rho >
    +\sqrt{\delta}$. Here, $\bm{x} ,\bm{y}$ are unit length vectors
    with inner product equal to $\rho$. The unit vector perpendicular
    $\bm{y}$ is given by $-(\bm{x}-\rho\bm{y})/\sqrt{1 - \rho^2}$. The
    circle $C$ is the projection of the feasible distortion sphere
    onto the $\bm{x}$-$\bm{y}$ plane. The problem in~\eqref{eq:PU_RVb} can be stated as finding $\bm{z}$ within the
    feasible distortion sphere which minimizes the cosine of the angle
    between $\bm{z}$ and $\bm{x}$. The optimum $\bm{z}$ lies on the
    tangent from the origin $O$ to the circle $C$ on the far side of
    $\bm{x}$. The dotted lines with hollow arrowheads denote the
    lengths of various quantities.}\label{fig:FD_case2}
\end{figure}
We can interpret the solution geometrically as shown in
Figure~\ref{fig:FD_case2}. Unlike the previous case ($0 \leq \rho \leq
+\sqrt{\delta}$), here the feasible distortion sphere does not allow
$\bm{z}$ to be perpendicular to $\bm{x}$. However, one can see that
the solution must lie on a tangent from the origin to the distortion
sphere. The optimum $\bm{z}$ in this case (Eq.~\eqref{eq:bestZ}) is
the addition of two vectors, one along $\bm{y}$ and the other
perpendicular to $\bm{y}$ (the unit vector along which is
$-(\bm{x}-\rho\bm{y})/\sqrt{1 - \rho^2}$). The coefficients for the
linear combination can be inferred from the geometry of the figure.
\end{proof}

%%%%%%%%%%%%%%%%%%%%%%%%%%%%%%%%%%%

%% file: side_info.tex
\section{\edits{Handling side information about sensitive  $X$}}\label{sec:side_info}
\edits{If there is some side information $U$ available about $X$, then certainly we cannot have less information leakage than $I(X;U)$. However if $U$ is a noisy transformation of $X$ independent of the randomness used by the PPAN mechanism in generating $Z$, we can still control any \emph{additional} information loss in the following manner. Using the properties of mutual information, we have that
\begin{equation*}%\label{eq:side_info}
I(X;Z,U) = I(X;Z) + I(X;U|Z) \leq I(X;Z) + I(X;U),
\end{equation*}
where the inequality above is true because $Z$ and $U$ are independent given $X$. So while $I(X;U)$ is already leaked, we can prevent further leakage by minimizing $I(X;Z)$.}

%% file: MI_utility.tex
\section{\edits{Mutual Information Utility}}
\label{sec:MI-utility}

While we focused on expected distortion to measure (dis)utility, our framework can be adapted to other general utility measures, for example, mutual information between the useful information and the released data.
The conditional entropy $h(Y|Z)$ is an alternative measure for distortion, which corresponds to the utility objective of maximizing the mutual information $I(Y;Z)$, since $h(Y)$ is fixed.
When $h(Y|Z)$ is used as the distortion measure in a scenario where the observation $W = X$, the privacy-utility tradeoff optimization problem, as described in Section~\ref{sec:formulation}, becomes equivalent to the \emph{Information Bottleneck} problem considered in~\cite{TishbyPB-Allerton00-InfoBottleneck}.
In other scenarios where the observation $W = Y$, this problem becomes the \emph{Privacy Funnel} problem introduced by~\cite{MakhdoumiSFW-ITW14-PrivacyFunnel}.
The formulation of~\eqref{eq:min-max} can be modified to address conditional entropy distortion by introducing another variational posterior $Q_{Y|Z}$ and using the following optimization, which applies a second variational approximation of mutual information,
\begin{align*}
\min_{P_{Z|W}, Q_{Y|Z}} \max_{Q_{X|Z}}\mathbb{E}\big[\log Q_{X|Z}(X|Z)\big] - \lambda \mathbb{E}\big[\log Q_{Y|Z}(Y|Z)\big],
\end{align*}
where the expectations are with respect to $(W,X,Y,Z) \sim P_{W,X,Y}
P_{Z|W}$, and the parameter $\lambda > 0$ can be adjusted to obtain
various points along the optimal tradeoff curve.
In a similar fashion to the approach in Section~\ref{sec:CorePPAN}, this optimization problem can be practically addressed via the training of three neural networks, which respectively parameterize the mechanism $P_{Z|W}$ and the two variational posteriors $Q_{X|Z}$ and $Q_{Y|Z}$.

%% file: surrogate.tex
\section{\edits{Handling vaguely-defined sensitive  $X$}}\label{sec:surrogate}
\edits{If the sensitive attribute is not well-defined, but there is an available surrogate $S$ for it such that the Markov chain
\begin{equation*}
X \leftrightarrow S \leftrightarrow Z
\end{equation*}
holds then $I(X;Z) \leq I(S;Z)$ by the data processing inequality.  Hence minimizing $I(S;Z)$ still gives us protection for the sensitive variable $X$. If such a Markov chain is not satisfied, i.e., $I(X;Z|S) \neq 0$ but $S$ is still a weak surrogate such that $I(X;Z|S) \leq \epsilon$ for some small $\epsilon > 0$, then since
\begin{equation*}%\label{eq:surrogate}
I(Z;X) \leq I(Z;X,S) = I(Z;S) + I(Z;X|S) \leq I(Z;S) + \epsilon,
\end{equation*}
minimizing $I(Z;S)$ still gives us some protection for $X$.}

\edits{Alternatively, if we cannot specify what attribute in the dataset $W$ is sensitive, we can still train PPAN by setting the sensitive attribute $X = W$, i.e., all attributes are regarded as sensitive. PPAN will then try to minimize information leakage for all attributes, while subject to the allowed distortion budget with respect to the useful attribute $Y$.} 

%% file: GMM.tex
\section{\edits{Sampling from a Gaussian Mixture Model (GMM)}}\label{sec:GMM}

The technique of sampling from a multivariate Gaussian described in Section~\ref{subsec:gaussian_release} can be extended to GMMs as follows. The release
mechanism can be realized with a neural network $f_\theta$ that
produces the set of parameters $\{(\boldsymbol{\mu}_{l,i},
\mathbf{A}_{l,i}, \pi_{l,i})\}_{l=1}^m = f_\theta(w_i)$, where $\pi_{l,i}$
are the mixture weights.  We then
sample $z_{l,i} = \mathbf{A}_{l,i}\mathbf{u}_{l,i} +
\boldsymbol{\mu}_{l,i}$ for each component distribution of the GMM,
and compute the loss terms via
\begin{equation} \label{eqn:gmm-loss}
\mathcal{L}^i_{\text{GMM}}(\theta, \phi) := \sum_{l=1}^m \pi_{l,i}
\big( \log Q_\phi(x_i|z_{l,i}) + \lambda d(y_i,z_{l,i}) \big),
\end{equation}
which combines the Gaussian sampling reparameterization trick with a
direct expectation over the mixture component selection.

%% file: discrete_baseline.tex
\section{\edits{Theoretically Optimal Privacy-Utility Tradeoffs for Symmetric Pair Distribution}}\label{sec:symm-pair-results}

The mutual information of the symmetric pair distribution (see \eqref{eq:symmetric_pmf1}) is given by~\cite{wang2017privacy} as
\begin{equation*}
I(X;Y) = \log m - p \log(m-1) - h_2(p) =: r_m(p),
\end{equation*}
where $h_2(p) := -p \log p - (1-p) \log (1-p)$ is the binary entropy function, and for convenience in later discussion, we define $r_m(p)$ as a function of the distribution parameters $m$ and $p$.

For sensitive and useful attributes jointly distributed according to the symmetric pair distribution, the theoretically optimal privacy-utility tradeoffs, as defined by~\eqref{eq:pp}, are analytically derived in~\cite{wang2017privacy} for several data observation models, while using probability of error as the distortion metric, i.e.,  $\mathbb{E}[\mathbf{1}(Y \neq Z)] = \Pr[Y \neq Z]$.
In one case, when the observation is the full data, i.e., $W = (X,Y)$, the optimal mutual information privacy-leakage as a function of the distortion (probability of error) limit $\delta \in [0,1]$ is given by
\begin{align} \label{eqn:SP_FD_region}
I^*_{W=(X,Y)}(\delta) =
\begin{cases}
r_m(p + \delta), & \text{if } \delta \leq 1-\frac{1}{m} - p, \\
r_m(p - \delta), & \text{if } \delta \leq p - (1 - \frac{1}{m}), \\
0, & \text{otherwise}.
\end{cases}
\end{align}
In another case, when the observation is only the useful attribute, i.e., $W = Y$, the optimal privacy-leakage as a function $\delta \in [0,1]$ is given by
\begin{align}\label{eqn:SP_OP_region}
I^*_{W=Y}(\delta) =
\begin{cases}
r_m \left( p + \delta \left(1 - \frac{pm}{m-1} \right) \right), & \text{if } \delta < 1-\frac{1}{m},\\
0, & \text{otherwise}.
\end{cases}
\end{align}

%% file: scatter_viz.tex
\section{\edits{Scatter plots of multivariate Gaussian data}}
\label{sec:multivar_Gdata_viz}
\begin{figure*}[!t]
\centering
\includegraphics[width=0.75\textwidth]{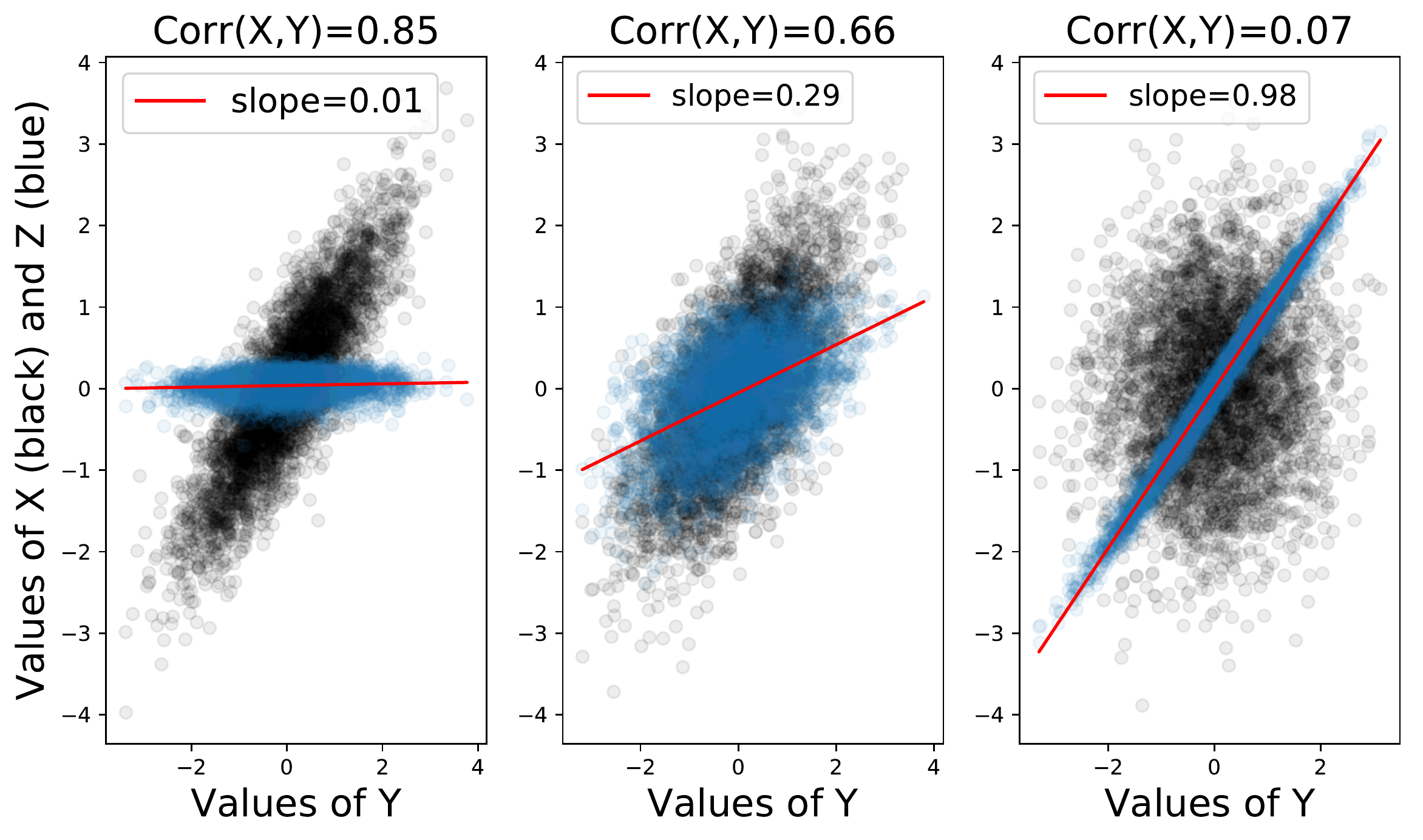}
\caption{Scatter plot of 3 components of 4000 test data points as input to a PPAN mechanism trained for $\delta=0.23$. The simulated data model is described in Section~\ref{sec:G_OPvec}. The black points are $(Y_i,X_i)$ test data component pairs and the blue points are $(Y_i, Z_i)$ pairs corresponding to the release generated by PPAN. The red line is a least squares fit of $Z_i$ v.s.\ $Y_i$ in each case. On components where $\text{Corr}(X_i, Y_i)$ is higher, PPAN mechanism allows higher levels of distortion, i.e., $(Y_i - Z_i)^2$. In addition to the different slope, the variance of the noise added in the three cases is also different.}
\label{fig:scatter_viz}
\end{figure*}
We visualize the output of the trained PPAN mechanism of Section~\ref{sec:G_OPvec} for a particular value of mean squared error. As shown in Figure~\ref{fig:vector_OP}, the operating points of the PPAN mechanism for various values of distortion $\delta$ is close to the theoretically optimal privacy-utility tradeoff. We focus on the PPAN mechanism trained corresponding to the operating point of $(I(X;Z), \mathbb{E}\lVert Y - Z\rVert^2) = (1.838, 0.230)$. 
Recall that the sensitive $X$, useful $Y$ (which is the same as the input $W$ to PPAN in the output perturbation observation model) and the released $Z$ are all vectors in $\mathbb{R}^5$. The distortion between $Y$ and $Z$ is measured as their component-wise squared error. The test set consists of 4000 $(X,Y)$ pairs, and we obtain 4000 realizations of $Z$ corresponding to the test set at the output of the trained PPAN. In Figure~\ref{fig:scatter_viz}, we show a scatter plot of three components of the test set and release for the chosen operating point. The three panels, each showing a particular component, illustrate the nature of noise added by PPAN in three different scenarios of high, medium and low correlation 
between the sensitive $X_i$ and useful $Y_i$. It can be seen that when the sensitive and useful variables are highly correlated, PPAN adds a significant amount of noise to obtain the release. On the other hand, when the correlation is low, PPAN adds very little noise and releases the useful $Y_i$ as is. The behavior for medium correlation is between the two extremes.